\documentclass[aps,twocolumn,prl,preprintnumbers,showpacs,amsmath,amssymb,superscriptaddress,longbibliography]{revtex4-1}
\usepackage[pdftex, colorlinks, citecolor=blue]{hyperref}   % this is for pdflatex pdf format
\usepackage{graphicx}
\usepackage{dcolumn}
\usepackage{threeparttable}
\usepackage{multirow}
\usepackage{booktabs}
\usepackage{txfonts}
\usepackage{xcolor}
\usepackage{bm}
\usepackage{amssymb}
\usepackage{amsmath}
\usepackage{latexsym}
\usepackage{epsfig}
\usepackage{amsbsy}
\usepackage{array}
\usepackage{tabularx}
\usepackage{esvect} %vector
\usepackage{extarrows}
\usepackage{float}
\usepackage[british]{babel}

\begin{document}

\title{Quantum Delocalization Enables Water Dissociation on Ru(0001)}

\author{Yu Cao}
\thanks{These authors contribute equally to this work.}
\affiliation{Shenyang National Laboratory for Materials Science, Institute of Metal Research, Chinese Academy of Sciences, 110016 Shenyang, China}
\affiliation{School of Materials Science and Engineering, University of Science and Technology of China, 110016 Shenyang, China}

\author{Jiantao Wang}
\thanks{These authors contribute equally to this work.}
\affiliation{Shenyang National Laboratory for Materials Science, Institute of Metal Research, Chinese Academy of Sciences, 110016 Shenyang, China}
\affiliation{School of Materials Science and Engineering, University of Science and Technology of China, 110016 Shenyang, China}

\author{Mingfeng Liu}
\affiliation{Shenyang National Laboratory for Materials Science, Institute of Metal Research, Chinese Academy of Sciences, 110016 Shenyang, China}

\author{Yan Liu}
\affiliation{Shenyang National Laboratory for Materials Science, Institute of Metal Research, Chinese Academy of Sciences, 110016 Shenyang, China}
\affiliation{School of Materials Science and Engineering, University of Science and Technology of China, 110016 Shenyang, China}

\author{Hui Ma}
\affiliation{Shenyang National Laboratory for Materials Science, Institute of Metal Research, Chinese Academy of Sciences, 110016 Shenyang, China}

\author{\\Cesare Franchini}
\affiliation{University of Vienna, Faculty of Physics and Center for Computational Materials Science, Kolingasse 14-16, A-1090 Vienna, Austria}
\affiliation{Dipartimento di Fisica e Astronomia, Universit\`{a} di Bologna, 40127 Bologna, Italy}

\author{Yan Sun}
\affiliation{Shenyang National Laboratory for Materials Science, Institute of Metal Research, Chinese Academy of Sciences, 110016 Shenyang, China}

\author{Georg Kresse}
\affiliation{University of Vienna, Faculty of Physics and Center for Computational Materials Science, Kolingasse 14-16, A-1090 Vienna, Austria}

\author{Xing-Qiu Chen}
\email{xingqiu.chen@imr.ac.cn}
\affiliation{Shenyang National Laboratory for Materials Science, Institute of Metal Research, Chinese Academy of Sciences, 110016 Shenyang, China}

\author{Peitao Liu}
\email{ptliu@imr.ac.cn}
\affiliation{Shenyang National Laboratory for Materials Science, Institute of Metal Research, Chinese Academy of Sciences, 110016 Shenyang, China}

\begin{abstract}
We revisit the long-standing question of whether water molecules dissociate on the Ru(0001) surface
through nanosecond-scale path-integral molecular dynamics simulations on a sizable supercell.
This is made possible through the development of an efficient and reliable machine-learning potential with near first-principles accuracy,
overcoming the limitations of previous \emph{ab initio} studies.
We show that the quantum delocalization associated with nuclear quantum effects
enables rapid and frequent proton transfers between water molecules, thereby facilitating the water dissociation on Ru(0001).
This work provides the direct theoretical evidence of water dissociation on Ru(0001),
resolving the enduring issue in surface sciences and offering crucial atomistic insights into water-metal interfaces.
\end{abstract}

\maketitle

The interaction of water with solid surfaces plays an important role in various scientific and technical fields,
including ice nucleation, catalysis, lubrication, and corrosion~\cite{C6CS00864J-2017,Knopf2023,MengshengBook2023}.
The subtle interplay between the water-solid interface and water-water hydrogen bonding interactions
results in complex structural motifs of water on surfaces, prompting extensive
experimental and theoretical investigations~\cite{HODGSON2009381,carrasco2012molecular,maier2015does,
bjorneholm2016water,shimizu2018water,zhou2021review,tian2022visualizing,GAO2024136}.
Among these studies, the water overlayer on the Ru(0001) surface is notably one of the most debated systems.

Initially, it was commonly believed that the first-layer water molecules on Ru(0001) would arrange themselves
into a bulk-ice-like bilayer~\cite{PhysRevLett.49.501,doering1982adsorption,10.1063/1.446561,PIRUG1991289}.
However, this notion was challenged by the low-energy electron diffraction (LEED) analysis
by Held and Menzel~\cite{HELD199492,doi:10.1142/S0218625X03005086},
which revealed nearly coplanar O atoms
with a vertical distance of just 0.14 $\AA$~\cite{doi:10.1142/S0218625X03005086},
significantly smaller than that of the ice-like bilayer.
Moreover, density functional theory (DFT) calculations
employing commonly used generalized gradient approximation functionals
predicted that the ice-like bilayer did not even wet the surface~\cite{feibelman2002partial}.
To resolve this discrepancy, Feibelman proposed a partially dissociated overlayer model~\cite{feibelman2002partial},
which not only reproduced the experimental observation of almost coplanar O atoms,
but also demonstrated superior energetic stability
compared to the H-up or H-down bilayers as well as the sublimation energy of ice-Ih~\cite{feibelman2002partial}.

Feibelman's study~\cite{feibelman2002partial} raised an even today debated question---whether
the water molecules within the overlayer dissociate on Ru(0001)?
Experimental evidence from sum-frequency generation vibrational spectroscopy~\cite{DENZLER2003618}
and reflection adsorption infrared spectroscopy~\cite{CLAY200489,PhysRevB.73.115414}
indicated that the water remained intact on Ru(0001) at low temperatures.
However, temperature programmed desorption measurements
observed residual atomic hydrogen on the surface after water desorption,
implying the dissociative adsorption~\cite{DENZLER2003113}.
The situation became even more puzzling by conflicting x-ray photoelectron spectroscopy (XPS)
studies~\cite{PhysRevLett.93.196101,FARADZHEV2005165,2008JCP,PhysRevLett.93.196102}.
Andersson \emph{et al.} attributed the flat water-hydroxyl overlayer observed by LEED to electron beam damage
and claimed that the water dissociation was a thermal or electron stimulated process~\cite{PhysRevLett.93.196101},
which was later supported by two independent XPS studies~\cite{FARADZHEV2005165,2008JCP}.
In contrast, Weissenrieder \emph{et al.} provided experimental XPS evidence for a partially dissociated water bilayer on Ru(0001)
at temperatures as low as 105 K using beam-damage-free low photon flux~\cite{PhysRevLett.93.196102}.
Besides XPS, scanning tunneling microscopy (STM) also unveiled perplexing results.
Tatarkhanov \emph{et al.}~\cite{doi:10.1021/ja907468m} and Maier \emph{et al.}~\cite{PhysRevB.85.155434,PhysRevLett.112.126101}
demonstrated through STM that at temperatures below 130 K, the deposited water remained intact and tended to form stripes of hexagons.
However, a more recent STM investigation by Schilling and Behm~\cite{schilling2018partial} indicated that
partial water dissociation could possibly occur already during slow adsorption at 120 K.

\begin{figure*}
\begin{center}
\includegraphics[width=0.9\textwidth,trim = {0.0cm 0.0cm 0.0cm 0.0cm}, clip]{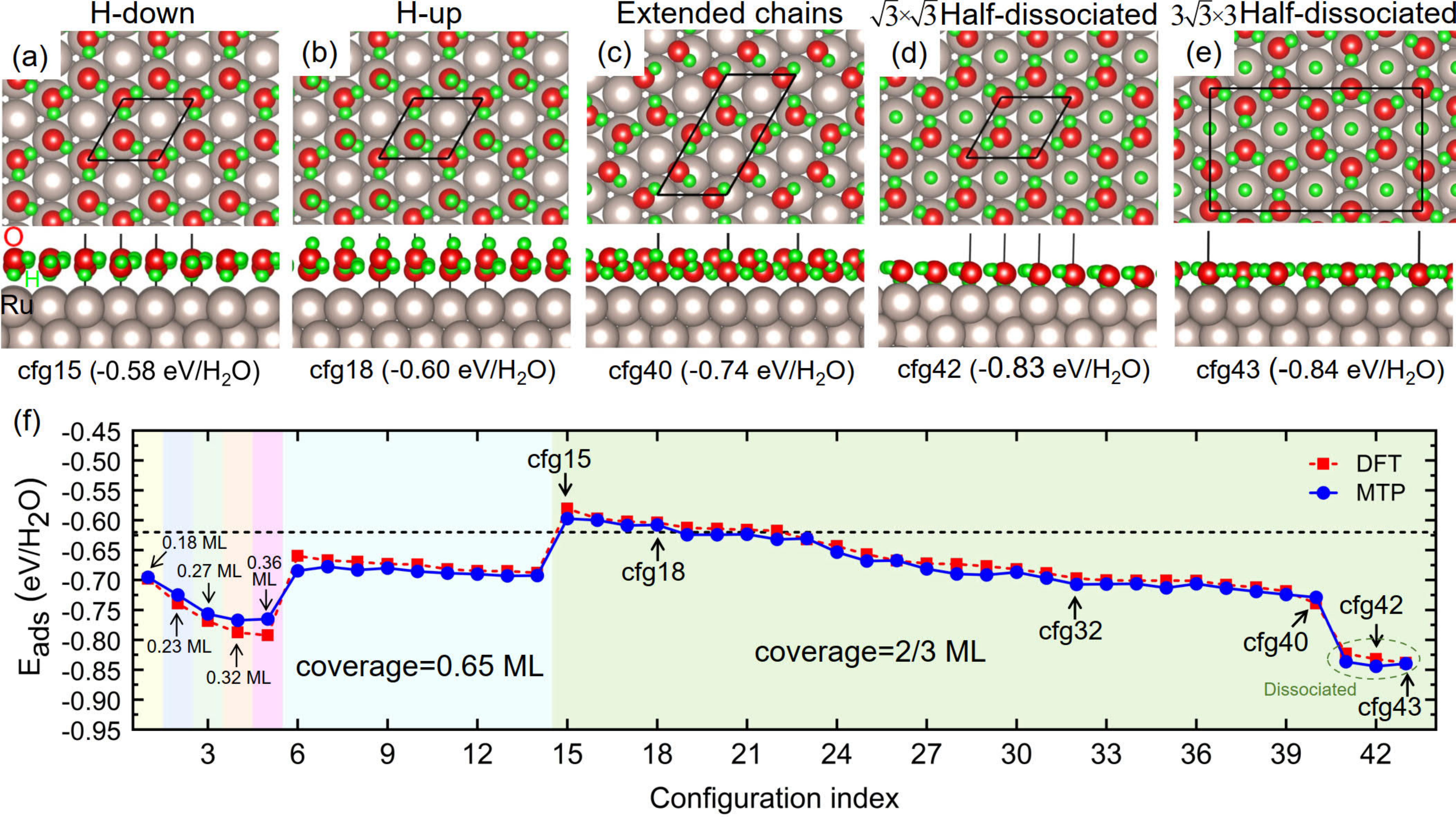}
\end{center}
\caption{Top and side views of (a) H-down, (b) H-up, (c) extended chains,
(d) $\sqrt{3}{\times}\sqrt{3}$ half-dissociated~\cite{PhysRevLett.93.196102}, and (e) $3\sqrt{3}\times3$ half-dissociated~\cite{feibelman2002partial} water overlayer models.
The black lines indicate the unit cell.
(f) The DFT and MTP predicted adsorption energies of H$_2$O on Ru(0001) for
the 43 considered configurations at various coverages (see Supplemental Material Fig.~S2~\cite{SM}).
The dashed line denotes the sublimation energy of ice-Ih.
}
\label{fig1}
\end{figure*}

The conflicting experimental results also stimulated extensive DFT calculations~\cite{michaelides2003different,PhysRevB.69.113404,
PhysRevB.69.195404,meng2005consistent,
PhysRevLett.90.216102,FEIBELMAN2005120,gallagher2009order,PhysRevLett.106.026101,herron2013atomic}.
Consistent with Feibelman's predictions, Michaelides \emph{et al.}~\cite{michaelides2003different,PhysRevB.69.113404}
confirmed that the partially dissociated overlayer was thermodynamically more stable than any identified water bilayers.
Nevertheless, the predicted dissociation energy barrier
was higher than the mean adsorption energy of the water bilayers~\cite{PhysRevB.69.195404,meng2005consistent},
indicating the favorable desorption of water over dissociation.
To address the problems associated with the coplanar overlayer observed by LEED
and the underestimated work function change upon adsorption for the half-dissociated model,
a metastable structure with mixed intact H-up and H-down molecules~\cite{meng2005consistent},
and an extended chains model containing chains of flat-lying and chains of H-down molecules~\cite{PhysRevB.73.115414,gallagher2009order}
were suggested.

Despite comprehensive DFT studies, the ground-state structure of water overlayers on Ru(0001) remains inconclusive
due to the vast configuration space originating from the complex hydrogen bonding networks and water-metal interactions.
Additionally, whether the water overlayers dissociate on Ru(0001) remains elusive,
given the high predicted dissociation energy barrier (0.5 eV~\cite{michaelides2003different} and 0.62 eV~\cite{meng2005consistent}).
Aiming to address these questions, we developed an accurate, efficient, and robust
machine-learning potential,
which not only facilitates the accurate exploration of the potential energy surface (PES) landscape for water adsorption on Ru(0001) at various coverages,
but also allows for long-time scale path-integral molecular dynamics (PIMD) simulations.
Our work provides direct and compelling evidence of water dissociation from an intact overlayer on Ru(0001),
which is driven by nuclear quantum effects (NQEs).

We begin by constructing and validating our moment tensor potential (MTP)~\cite{Alexander2016}.
Following the scheme outlined in Refs.~\cite{PhysRevLett.130.078001,Liu2024},
the training dataset was initially acquired through an on-the-fly active learning procedure~\cite{JinnouchiPRL2019,JinnouchiPRB2019}
implemented in the Vienna \emph{ab initio} simulation package (VASP)~\cite{PhysRevB.47.558,PhysRevB.54.11169},
and subsequently expanded iteratively using the active learning method implemented in the MLIP package~\cite{Novikov_2021}.
First-principles calculations were performed using VASP~\cite{PhysRevB.47.558,PhysRevB.54.11169}
employing the revised Perdew-Burke-Ernzerhof (RPBE) functional for its improved adsorption energetics~\cite{PhysRevB.59.7413}.
Van der Waals (vdW) interactions were included using the D3 method of Grimme \emph{et al.}~\cite{10.1063/1.3382344} with zero damping.
We found that RPBE+D3 yielded similar Ru-O and O-O vertical distances and relative water adsorption energies
when compared to SCAN~\cite{PhysRevLett.115.036402},
PBE+D3~\cite{PhysRevLett.77.3865,10.1063/1.3382344}, and optB88-vdW~\cite{klimevs2009chemical} functionals (Supplemental Material Table~S1 and Fig.~S1~\cite{SM}).
The training dataset contained 6,813 structures including bulk ice-Ih, bulk Ru, and the clean Ru(0001) surface,
as well as intact and dissociated water overlayers at various coverages up to 2/3 monolayer (ML) (Supplemental Material Table~S2~\cite{SM}).
To enhance the efficiency, the MTP was fitted with the optimized basis sets
by refining the contraction process of moment tensors using our in-house code (IMR-MLP)~\cite{wang2024}.
The MTP was validated on a test dataset of 600 structures including bulk Ru, the clean Ru(0001) surface,
intact and half-dissociated water overlayers at 2/3 ML (Supplemental Material Table~S3~\cite{SM}).
These test structures were randomly sampled from MD simulations.
The MTP demonstrates a high accuracy, with validation root-mean-square errors of 1.02 meV/atom for energies
and 59 meV/$\AA$ for forces (Supplemental Material Fig.~S3~\cite{SM}).
More computational details can be found in the Supplemental Material~\cite{SM}.

\begin{figure*}
\begin{center}
\includegraphics[width=0.99\textwidth,trim = {0.0cm 0.0cm 0.0cm 0.0cm}, clip]{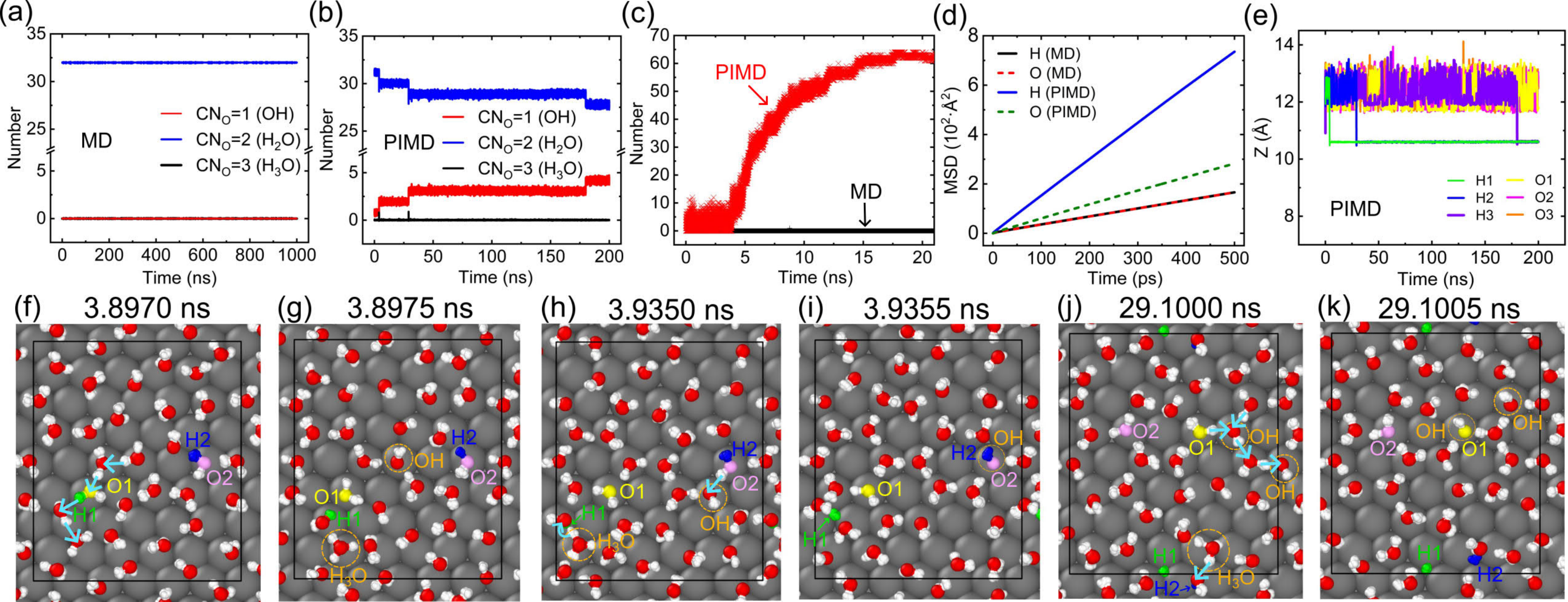}
\end{center}
\caption{Time evolution of the number of OH (CN$_{\rm O}$=1),  H$_2$O (CN$_{\rm O}$=2), and H$_3$O (CN$_{\rm O}$=3) species
in (a) classical MD and (b) PIMD simulations at 300 K. CN$_{\rm O}$ represents the coordination number of the oxygen atom.
(c) Time evolution of occurrences when either of the two hydrogen atoms in an H$_2$O molecule
differs from the initial configuration at 0 ns.
(d) Mean square displacements (MSD) of the O and H atoms.
(e) Time evolution of the vertical distances of the H and O atoms [specified in (f)]  in the PIMD simulation.
(f)-(k) Snapshots from the PIMD trajectory. The tracked O and H atoms as well as the H$_3$O and OH species are highlighted.
The cyan arrows depict the proton transfer process.
}
\label{fig2}
\end{figure*}

The developed MTP allows us to efficiently explore the PES of water overlayers at different coverages.
As depicted in Fig.~\ref{fig1}, for the considered 43 configurations including intact and dissociated water overlayers,
the MTP accurately predicts their adsorption energies, showing remarkable agreement with DFT results.
Particularly, we focus on the extensively studied 2/3 ML overlayers.
It is evident that the H-down and H-up  models exhibit comparable adsorption energies.
However, both configurations would not wet the Ru(0001) surface,
since their adsorption energies are lower than the sublimation energy of ice-Ih
[see Fig.~\ref{fig1}(f) and Supplemental Material Fig.~S1(a)~\cite{SM}].
The most stable intact overlayer identified at 2/3 ML is the extended chains model with a unit cell of $\sqrt{3}{\times}2\sqrt{3}$,
where the O atoms in the flat-lying chains are positioned atop Ru, whereas the O atoms in the H-down chains occupy the bridge sites [see Fig.~\ref{fig1}(c)].
Note that this extended chains model is more stable by 0.04 eV/H$_2$O
than the one originally proposed by Hodgson \emph{et al.} (i.e.,  cfg32)~\cite{PhysRevB.73.115414,gallagher2009order}, where all the O atoms are situated atop Ru.
Generally, the dissociated overlayers (configurations 41-43~\cite{feibelman2002partial,PhysRevLett.93.196102})
exhibit coplanar O atoms and are more stable compared to the intact ones,
consistent with previous calculations~\cite{feibelman2002partial,michaelides2003different,PhysRevB.69.113404,PhysRevB.69.195404,meng2005consistent}.

We note that the overlayer configurations examined above are constrained to ordered phases with small unit cells,
without fully incorporating the disorder effects.
Experiments also face challenges in identifying hydrogen disorder due to its light mass~\cite{PhysRevLett.87.216102}.
To examine the disorder effects, we performed simulated annealing (SA) MD simulations
on large supercells using the MTP and the LAMMPS code~\cite{Thompson2022}.
The most favorable intact (cfg40) and dissociative (cfg43) configurations
were selected as initial structures.
Two large supercells (i.e., $6{\times}3\sqrt{3}$ and $12{\times}6\sqrt{3}$)
were employed to account for varying degrees of disorder.
For each configuration, 90 independent SA runs were conducted, followed by structural relaxations.
Interestingly, for the dissociative case, the lowest-energy configuration identified by SA
exhibits only marginally higher stability (0.01 eV/H$_2$O energy gain) in comparison to the initial ordered phase (Supplemental Material Fig.~S4~\cite{SM}).
The slightly higher stability of cfg43 compared to cfg42 is also such a manifestation of disorder effects (Fig.~\ref{fig1}).
The disordered configurations accommodated within large supercells
primarily exhibit disorder in hydrogen orientation, aimed at maximizing hydrogen bonding.
However, for the intact case, we were unable to find a disordered configuration more favorable than the
$\sqrt{3}{\times}2\sqrt{3}$ extended chains model within the 90 SA runs (Supplemental Material Fig.~S4~\cite{SM}).
This indicates a flat PES with numerous local minima, in agreement with Hodgson's findings~\cite{gallagher2009order}.
In fact, the extended chains configuration is a model
that already accounts for partial disorder in the O height and proton orientation~\cite{gallagher2009order}.
For both the dissociative and intact cases, a supercell beyond the $6{\times}3\sqrt{3}$ unit cell almost shows no energy gain from disorder.

Having identified the most stable intact and dissociative overlayers,
we now turn to tackling the enduring question of whether water overlayers dissociate on Ru(0001).
\textcolor{black}{First, we computed the dissociation energy barrier,
yielding 0.53 eV for dissociating half of the water molecules,
aligning with previous predictions~\cite{michaelides2003different,meng2005consistent}.
The energy barrier is found to be higher ($>$0.6 eV) for dissociating 1/8 of the water molecules (Supplemental Material Fig.~S5~\cite{SM})}.
Based on a rough estimation from transition state theory~\cite{doi:10.1021/jp953748q},
a dissociation energy barrier of this magnitude would necessitate a time scale exceeding microseconds,
far beyond the accessible time-scale of MD simulations.
Indeed, we did not observe the dissociation throughout a one-microsecond classical MD simulation [Fig.~\ref{fig2}(a)].
Nevertheless, classical MD simulations do not take into account the NQEs,
which have been shown to be significant in hydrogen-related systems~\cite{10.1063/1.471221,10.1038/32609,10.1038/17579,
PhysRevLett.100.166101,10.1063/1.3142828,LiXinZhengPRL2010,LixinzhengPNAS_2011,Chen2013,Lixinzheng_CPLReview_2015,Lixinzheng_Science2016,
Michele_Review2016,LiXinZheng_review_2017,Chen2018,Markland2018,XinZhengLi2019_review,
 Chenbingqing_PNAS2019,doi:10.1021/acs.jpclett.0c01025,Bore2023, doi:10.1126/science.ads4369}.
In particular,  Li \emph{et al.}~\cite{LiXinZhengPRL2010} showcased the pronounced NQEs
for water overlayers on transition metal surfaces through  \emph{ab initio} PIMD simulations.
Despite their high accuracy, \emph{ab initio} methods face limitations concerning the accessible time and length scales,
hindering the direct observation of water dissociations.

\begin{figure}
\begin{center}
\includegraphics[width=0.49\textwidth,trim = {0.0cm 0.0cm 0.0cm 0.0cm}, clip]{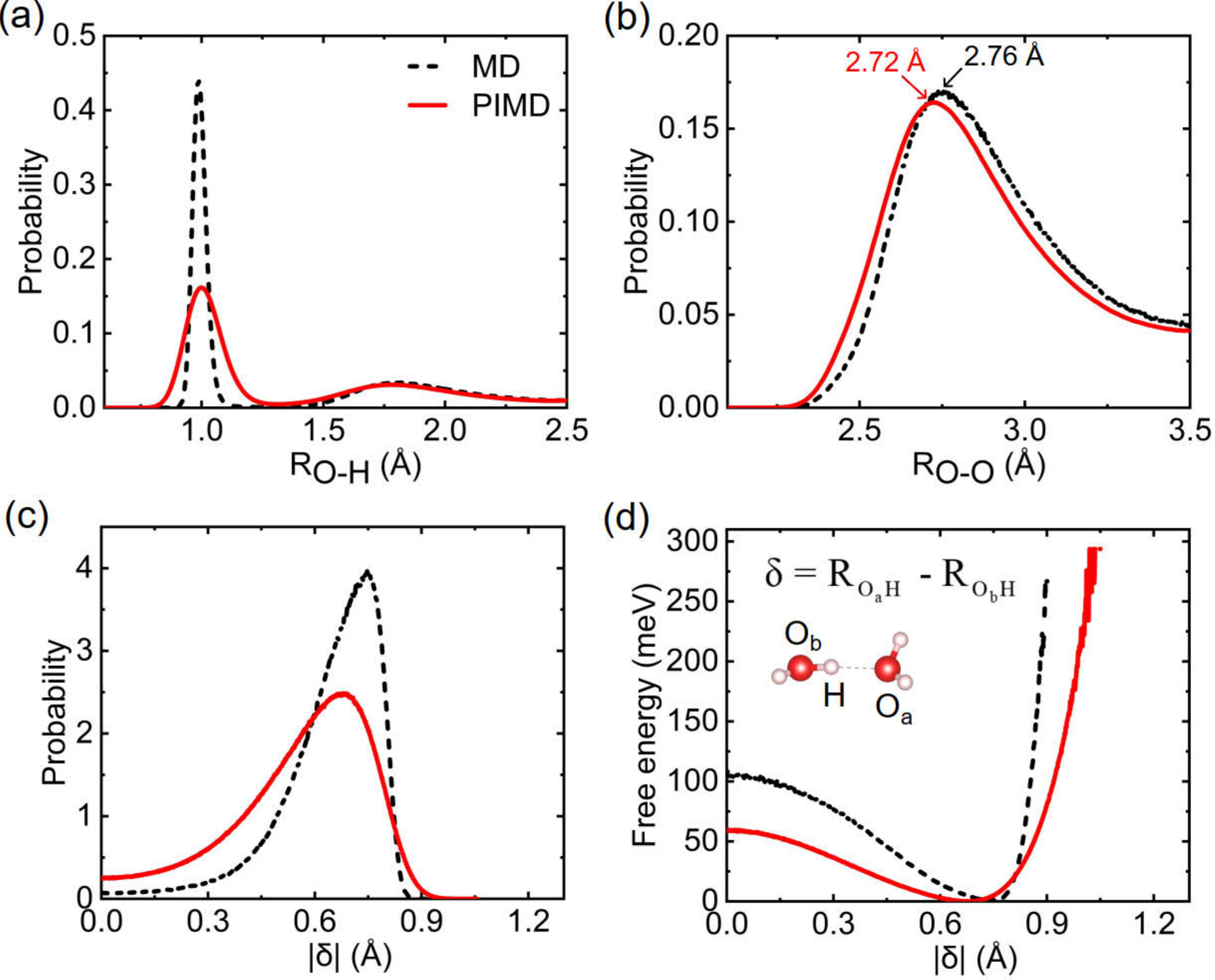}
\end{center}
\caption{Probability distributions of (a) O-H and (b) O-O distances at 300 K.
(c) Probability distribution of the proton transfer coordinate $|\delta|=|R_{\rm O_aH} - R_{\rm O_bH}|$,
where $R_{\rm O_aH}$ denotes the distance between the O$_{\rm a}$ atom and the H atom [see the insert in (d)].
$|\delta|$=0 indicates that the shared H atom is exactly positioned between two O atoms.
(d) Free energy profile for the protons along the intermolecular axes.
}
\label{fig3}
\end{figure}

Thanks to our accurate and efficient machine-learning potential, we were able to perform PIMD simulations on large supercells
for an extended timescale,  while maintaining \emph{ab initio} accuracy.
In the PIMD simulations, 16 beads were employed to sample the imaginary time path-integral and the statistical analyses were derived by averaging these 16 beads.
We employed a $6\times4\sqrt{3}$ supercell with 32 H$_{2}$O molecules based on the most stable intact overlayer configuration (i.e., extended chains model) as the initial structure.
Throughout a 200-nanosecond PIMD simulation,
three abrupt changes in the number of OH,  H$_2$O, and H$_3$O species are observed [see Fig.~\ref{fig2}(b)],
indicating three instances of water dissociation.
To corroborate our findings, we conducted additional independent PIMD and classical MD simulations.
Water dissociation was consistently reproduced in all PIMD simulations but was absent in classical MD simulations within the considered timescale (Supplemental Material Fig.~S7~\cite{SM}).
\textcolor{black}{Furthermore, PIMD simulations of D$_2$O adsorption on Ru(0001), conducted under identical computational settings as
H$_2$O adsorption, revealed no dissociation. This result underscores the pronounced kinetic isotope effect and aligns with experimental observations~\cite{CLAY200489}.}
Collectively, these results demonstrate that the NQEs facilitate the water dissociation on Ru(0001).

It is worth noting that such NQEs cannot be simply interpreted as
the commonly assumed thermalization effect in classical MD simulations at elevated temperatures.
This is because, even when the temperature is raised to 350 K,
water dissociation does not occur within the same time frame in the classical MD simulation  (Supplemental Material Fig.~S7~\cite{SM}).
Importantly, the NQEs result in a significant quantum delocalization of protons.
This is evidenced by the broadened first peak in the probability distributions of O-H distances [Fig.~\ref{fig3}(a)]
as well as the enhanced probability distributions
of protons within the central region of two adjacent oxygen atoms [Fig.~\ref{fig3}(c)]
observed in PIMD simulations, as compared to classical MD simulations.
Moreover, the quantum delocalization of the shared protons between the two oxygen atoms
slightly shortens the O-O distances [Fig.~\ref{fig3}(b)], strengthening the hydrogen bonding.
Our results agree well with the findings of Li \emph{et al.}~\cite{LiXinZhengPRL2010}.
Furthermore, we computed the free energy profile for the protons along the intermolecular axes
through $\Delta  F(|\delta|)=-k_BT\ln [P(|\delta|)]$ following Ref.~\cite{LiXinZhengPRL2010},
where $P(|\delta|)$ represents the probability distribution
of $|\delta|=|R_{\rm O_aH} - R_{\rm O_bH}|$ [see Fig.~\ref{fig3}(c)] and $k_B$ is the Boltzmann constant.
One can see from Fig.~\ref{fig3}(d) that the free energy barrier for proton transfer
is just 59 meV in the PIMD simulation, a value notably lower than that observed in the classical MD simulation ($\sim$103 meV).

The small free energy barrier in PIMD simulations enables rapid and frequent proton transfers among water molecules [Figs.~\ref{fig2}(c)-(d)].
Moreover, the emergence of water dissociation would enhance the rate of proton transfers [Fig.~\ref{fig2}(c)].
Conversely, in classical MD simulations, the protons almost do not undergo transfers due to the large free energy barrier [Fig.~\ref{fig2}(c)].
Specifically, in classical MD, the two H atoms and an O atom in an H$_2$O molecule consistently exhibit identical motions,
whereas in PIMD, they follow distinct motion pathways due to the proton transfers [Fig.~\ref{fig2}(d)].
Interestingly, we observe that while the dissociated H atoms diffuse most rapidly with a diffusion coefficient of 5.58$\times10^{-5}$ cm$^2$/s,
the diffusion of H and O atoms within the water-hydroxyl overlayer is comparatively slower than within the intact overlayer (Supplemental Material Fig.~S8~\cite{SM}).
This might be related to the inhibited proton transfer by hydroxide ions~\cite{Chen2018}.

To unveil the details of proton transfers and water dissociation, we analyzed the PIMD trajectory.
Note that although PIMD does not provide a real-time picture of the trajectories,
we still believe that the mechanisms we observe would prevail in a centroid molecular dynamics
and that the observed mechanisms are highly relevant to water dissociation.
Initially, ephemeral hydroxyls exist due to the quantum delocalization associated with the NQEs,
as evidenced by the fluctuation between the number of hydroxyls and intact H$_2$O (Supplemental Material Fig.~S9~\cite{SM}).
At 3.8975 ns, the formation of an H$_3$O and OH pair commences due to the proton transfer [Figs.~\ref{fig2}(f)-(g)],
and the pair would typically last for $\sim$38 ps (Supplemental Material Fig.~S9~\cite{SM}).
At 3.9355 ns, the first instance of water dissociation emerges.
Interestingly, we observe that the dissociation does not directly originate from H$_3$O, but instead undergoes an indirect process.
Specifically, one of the protons in H$_3$O is transferred to the neighboring H$_2$O,
leading to the simultaneous dissociation of a hydrogen atom from the latter [Figs.~\ref{fig2}(h)-(i)].
The dissociated H is observed to preferentially occupy the hollow fcc sites rather than the top sites on Ru(0001).
This observation is consistent with our DFT calculations, when the dissociation fraction is low [Supplemental Material Fig.~S5(l)~\cite{SM}].
Moreover, we find that once dissociated, the dissociated H decreases its vertical height [Fig.~\ref{fig2}(e)],
diffuses on the surface,  but is rarely involved in proton transfers.
The existence of the dissociated hydroxyl increases the frequency of proton transfers [Fig.~\ref{fig2}(c)].
At 29.1005 ns, the second occurrence of water dissociation takes place, following a procedure akin to the first instance [Figs.~\ref{fig2}(j)-(k) and Supplemental Material Fig.~S10~\cite{SM}].
\textcolor{black}{At 179.9685 ns, a third water molecule dissociates [Fig.~\ref{fig2}(b)].
Notably, this third dissociation event does not involve the formation of H$_3$O (Supplemental Material Fig.~S11~\cite{SM}).}
At this point, the system contains 29 intact water molecules, 3 hydroxyls, and 3 dissociated hydrogen atoms.
Over extended time scales, a greater number of H$_2$O would likely dissociate.
\textcolor{black}{Note that within the simulated time frame, no water desorption was observed
due to the higher energy barrier for desorption compared to dissociation (Supplemental Material Figs.~S6 and S5~\cite{SM}).
This aligns with experimental findings that H$_2$O dissociates prior to desorption~\cite{CLAY200489}.}
However, the large computational cost of PIMD kept us from much longer simulations.

In summary, we have developed an efficient and robust machine-learning potential
to accurately capture the intricate interactions between water overlayers and the Ru(0001) surface
through active learning and basis sets optimization.
This enables us to identify the most stable overlayers for both intact and dissociative adsorption
and facilitate the execution of PIMD simulations with near first-principles accuracy
on a time and length scale that is inaccessible through \emph{ab initio} methods.
This capability is crucial, as it allows us to directly observe water dissociation on Ru(0001)
through a nanosecond-scale PIMD simulation, a remarkable achievement that is unattainable with previous \emph{ab initio} studies.
We clearly demonstrate that the water dissociation is driven by the quantum delocalization originating from NQEs.
These are essential to facilitate rapid and frequent proton transfers.
Furthermore, the formation of H and OH pairs appears to occur via long-range density fluctuations,
with an H and OH pair forming quasi-instantly over a distance of up to 5 connected H$_2$O molecules.
This is facilitated by a combination of the Grotthuss mechanism and quantum fluctuations.
To our knowledge, such a long-range mechanism has not been observed before.
Our study implies that quantum fluctuations can dramatically change the barriers to water dissociation, but also lead to new dissociation pathways.
This has far-reaching implications not only for surface science, electrolysis and fuel cells, but also for biology, where water is ubiquitous.
Clearly, our study suggests that quantum fluctuations must be taken into account whenever water dissociation occurs.
The implications for the now so important field of hydrogen production are huge, as the barriers appear to be cut in half by quantum fluctuations.

% \section*{Acknowledgements}
This work is supported by
the National Natural Science Foundation of China (Grants No.~52422112, No.~52188101, and No.~52201030),
the Strategic Priority Research Program of the Chinese Academy of Sciences (XDA041040402),
the Liaoning Province Science and Technology Major Project (2024JH1/11700032, 2023021207-JH26/103 and RC230958),
the National Key R{\&}D Program of China 2021YFB3501503,
and
the Special Projects of the Central Government in Guidance of Local Science and Technology Development (2024010859-JH6/1006).

\emph{Data availability}---The data that support the findings of this article are openly available~\cite{DataShare}.

\bibliography{Ru-H2O-Reference}

\end{document}

\title{Supplemental Material to \\
``Quantum Delocalization Enables Water Dissociation on Ru(0001)"}

\author{Yu Cao}
\thanks{These authors contribute equally to this work.}
\affiliation{Shenyang National Laboratory for Materials Science, Institute of Metal Research, Chinese Academy of Sciences, 110016 Shenyang, China}
\affiliation{School of Materials Science and Engineering, University of Science and Technology of China, 110016 Shenyang, China}

\author{Jiantao Wang}
\thanks{These authors contribute equally to this work.}
\affiliation{Shenyang National Laboratory for Materials Science, Institute of Metal Research, Chinese Academy of Sciences, 110016 Shenyang, China}
\affiliation{School of Materials Science and Engineering, University of Science and Technology of China, 110016 Shenyang, China}

\author{Mingfeng Liu}
\affiliation{Shenyang National Laboratory for Materials Science, Institute of Metal Research, Chinese Academy of Sciences, 110016 Shenyang, China}

\author{Yan Liu}
\affiliation{Shenyang National Laboratory for Materials Science, Institute of Metal Research, Chinese Academy of Sciences, 110016 Shenyang, China}
\affiliation{School of Materials Science and Engineering, University of Science and Technology of China, 110016 Shenyang, China}

\author{Hui Ma}
\affiliation{Shenyang National Laboratory for Materials Science, Institute of Metal Research, Chinese Academy of Sciences, 110016 Shenyang, China}

\author{\\Cesare Franchini}
\affiliation{University of Vienna, Faculty of Physics and Center for Computational Materials Science, Kolingasse 14-16, A-1090 Vienna, Austria}
\affiliation{Dipartimento di Fisica e Astronomia, Universit\`{a} di Bologna, 40127 Bologna, Italy}

\author{Yan Sun}
\affiliation{Shenyang National Laboratory for Materials Science, Institute of Metal Research, Chinese Academy of Sciences, 110016 Shenyang, China}

\author{Georg Kresse}
\affiliation{University of Vienna, Faculty of Physics and Center for Computational Materials Science, Kolingasse 14-16, A-1090 Vienna, Austria}

\author{Xing-Qiu Chen}
\email{xingqiu.chen@imr.ac.cn}
\affiliation{Shenyang National Laboratory for Materials Science, Institute of Metal Research, Chinese Academy of Sciences, 110016 Shenyang, China}

\author{Peitao Liu}
\email{ptliu@imr.ac.cn}
\affiliation{Shenyang National Laboratory for Materials Science, Institute of Metal Research, Chinese Academy of Sciences, 110016 Shenyang, China}

\maketitle

%--------------------------------------------------------------------------------
\section{First-principles calculations}\label{sec:DFT_details}
%--------------------------------------------------------------------------------
The first-principles calculations were performed using the Vienna \emph{ab initio} simulation package (VASP)~\cite{PhysRevB.47.558, PhysRevB.54.11169}.
The generalized gradient approximation of revised Perdew-Burke-Ernzerhof functional (RPBE)~\cite{PhysRevB.59.7413} was used for the exchange-correlation functional~\cite{PhysRevB.59.7413}.
The projector augmented wave (PAW) pseudopotentials~\cite{PhysRevB.50.17953,PhysRevB.59.1758}  ({\tt Ru\_sv}, {\tt H} and {\tt O}) were adopted.
A plane wave cutoff of 450 eV and a $\Gamma$-centered $k$-point grid with a spacing of 0.201~$\AA^{-1}$ between $k$ points were employed.
The Gaussian smearing method with a smearing width of 0.05 eV was used.
The electronic optimization was performed
until the total energy difference between two consecutive  iterations was less than 10$^{-6}$ eV.
The structures were optimized until the forces were smaller than 0.02 eV/$\AA$.
The van der Waals (VdWs) interaction was accounted for via the D3 method of Grimme \emph{et al.}~\cite{10.1063/1.3382344} with standard zero damping.

The water adsorption energy on Ru(0001) (${E_{ads}}$, in eV/H$_2$O)  was calculated as
\begin{equation}\label{eq:E_ads}
{{E_{ads}} = ({E_{{\rm{total}}}} - {E_{{\rm{surf}}}} - {\rm{n }}E_{{\rm{{H_2}O}}})/n},
\end{equation}
where ${E_{{\rm{total}}}}$ is the total energy of the water overlayer on Ru(0001),
$E_{{\rm{surf}}}$ is the total energy of the clean surface, $n$ is the number of adsorbed H$_2$O molecules, and $E_{{\rm{{H_2}O}}}$ is the total energy of the H$_2$O molecule.
The slab was modeled using 5 layers with the bottom three layers fixed and a vacuum width of 15~$\AA$.

The sublimation energy of bulk ice-Ih (${E_{sub}}$, in eV/H$_2$O) was calculated as
\begin{equation}\label{eq:E_for}
{{E_{sub}} = ({E_{{\rm{ice-Ih}}}} - {\rm{n }}E_{{\rm{{H_2}O}}})/n},
\end{equation}
where ${E_{{\rm{ice-Ih}}}}$ is the total energy of bulk ice-Ih, $n$ is the number of H$_2$O formula units in the bulk ice-Ih, and $E_{{\rm{{H_2}O}}}$ is the total energy of the H$_2$O molecule.

The surface energy of Ru(0001) (${E_{sf}}$) was calculated as
\begin{equation}\label{eq:E_sur}
{{E_{sf}} = \frac{1}{{2A}}({E_{{\rm{slab}}}} - N{E_{{\rm{bulk}}}})},
\end{equation}
where ${E_{{\rm{slab}}}}$ is the total energy of the slab, ${E_{{\rm{bulk}}}}$ is the total energy per atom of bulk Ru,
$N$ is the number of Ru atoms in the slab, and $A$ is the surface area.

\newpage
\clearpage
%--------------------------------------------------------------------------------
\section{Assessment of exchange-correlation functionals}\label{sec:DFT_ex}
%--------------------------------------------------------------------------------

\begin{figure}
\begin{center}
\includegraphics[width=1.0\textwidth,trim = {0.0cm 0.0cm 0.0cm 0.0cm}, clip]{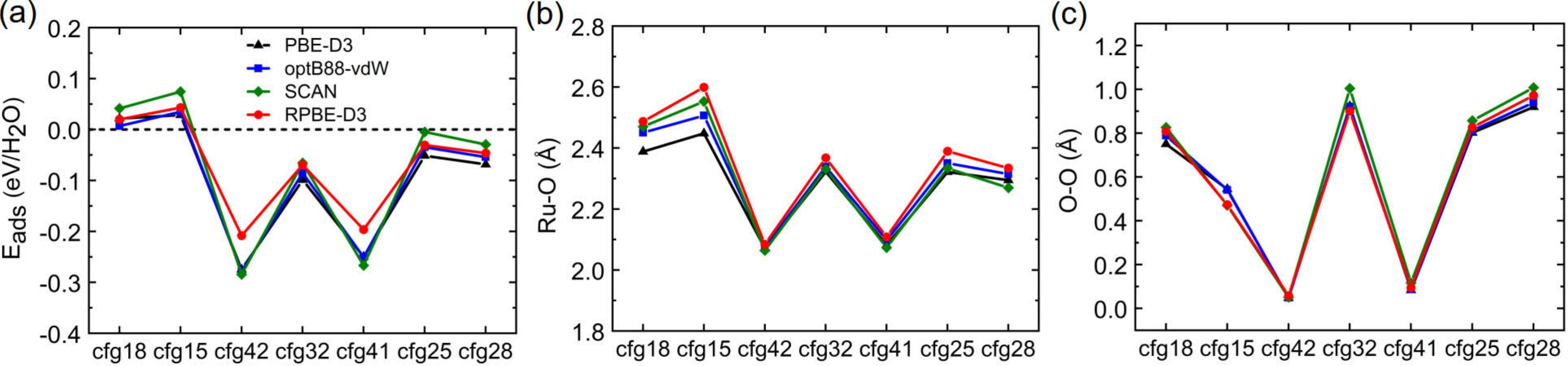}
\end{center}
\caption{(a) The relative energies of water adsorption with respect to the sublimation energy of ice-Ih predicted by four functionals.
(b) and (c) depict the vertical distances between Ru-O and O-O, respectively.}
\label{figs1_functional}
\end{figure}

\begin{figure}
\begin{center}
\includegraphics[width=0.85\textwidth,trim = {0.0cm 0.0cm 0.0cm 0.0cm}, clip]{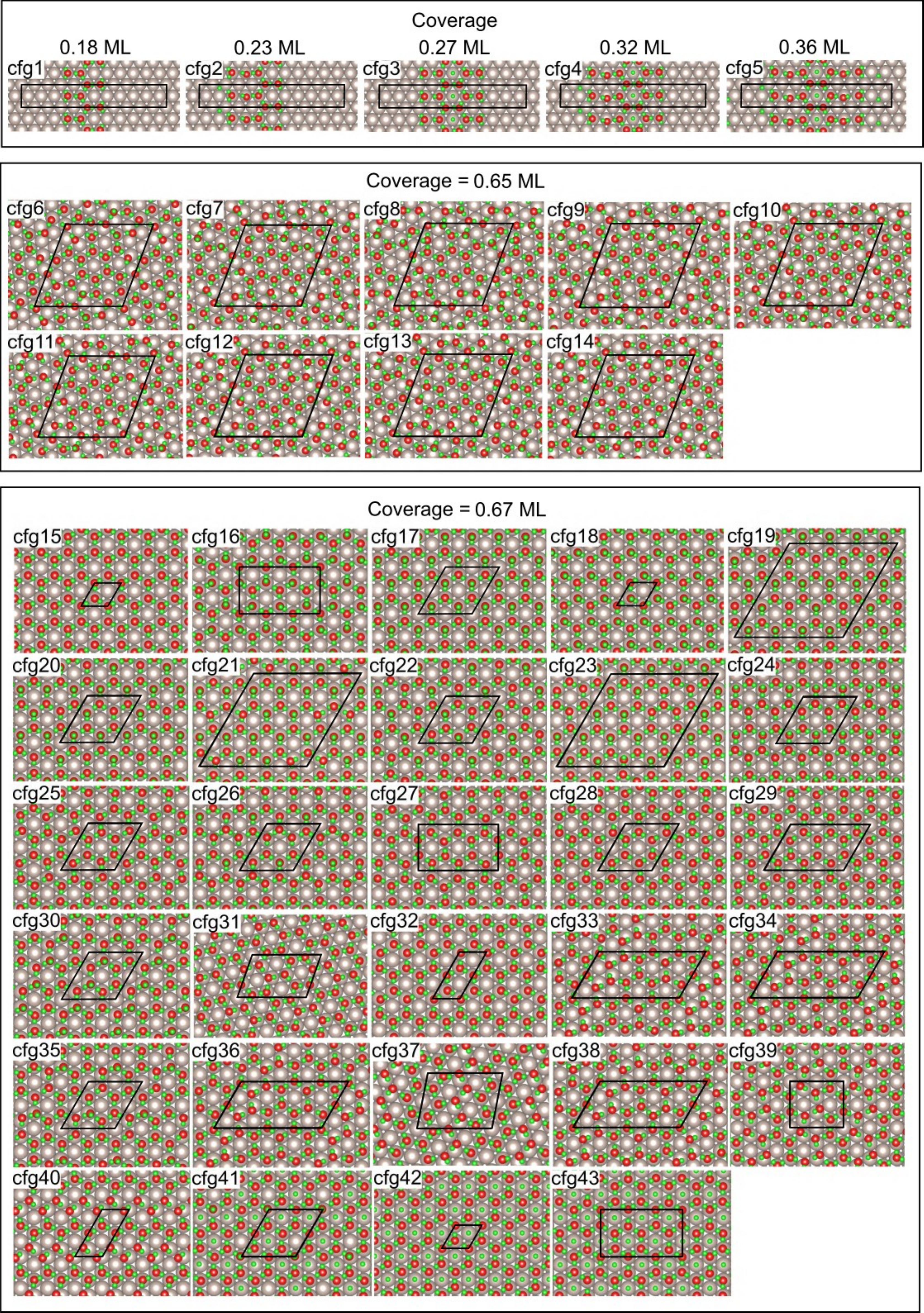}
\end{center}
\caption{The considered 43 water overlayer configurations on Ru(0001) with various coverages,
for which the adsorption energies were calculated and displayed in Figure 1 of the main text.
The configurations from cfg1 to cfg4 were take from Ref.~\cite{10.1063/1.2988903}.
The configuration cfg5 was extracted from Ref.~\cite{schilling2018partial,10.1063/1.2988903}.
The configurations of cfg6-cfg14, cfg17, cfg19-cfg24, cfg26, cfg30, and cfg35 were taken from Ref.~\cite{zhao2022intrinsic}.
The configurations of  cfg15, cfg16, cfg18, and cfg43 were extracted from Ref.~\cite{feibelman2002partial}.
The configuration of cfg32 was obtained from Ref.~\cite{PhysRevLett.106.026101}.
The configurations of cfg25, cfg27-cfg29, cfg31, cfg33-cfg34, and cfg36-cfg39 were derived from Ref.~\cite{gallagher2009order}.
The configurations of cfg41 and cfg42 were obtained from Ref.~\cite{PhysRevLett.93.196102}.
The Ru, O, and H atoms are represented by brown, red, and green spheres, respectively.
The black lines indicate the unit cell.
}
\label{figs2_structure}
\end{figure}

To assess the performance of the RPBE+D3 method~\cite{PhysRevB.59.7413,10.1063/1.3382344},
we computed the lattice parameters of bulk Ru, surface energy of Ru(0001), the sublimation energy of ice-Ih, and
the adsorption energy as well as the vertical distances between Ru-O and O-O of seven selected water overlayer configurations
including the H-up bilayer (cfg18), H-down bilayer (cfg15), half-dissociated (cfg42)~\cite{PhysRevLett.93.196102},
extended chains (cfg32)~\cite{gallagher2009order}, modified half-dissociated (cfg41)~\cite{PhysRevLett.93.196102},
two-linear-chains (cfg25)~\cite{gallagher2009order}, and four-arc-chains (cfg28)~\cite{gallagher2009order}.
The configurations were denoted according to Figure 1 in the main text.
Their crystal structures were illustrated in Fig.~\ref{figs2_structure}.
The RPBE+D3 computed results were compared with those predicted by
PBE+D3~\cite{PhysRevLett.77.3865}, optB88-vdW~\cite{klimevs2009chemical}, and SCAN~\cite{PhysRevLett.115.036402},
and summarized in Table~\ref{tables1}.
Furthermore, Figure~\ref{figs1_functional} depicted the vertical distances between Ru-O and O-O and water adsorption energies predicted by four functionals.
It is evident that the four functionals investigated yielded largely similar outcomes.
Because of the improved description for the adsorption energetics by RPBE~\cite{PhysRevB.59.7413}
and the excellent description of the water behavior by RPBE+D3~\cite{10.1063/1.4892400,doi:10.1073/pnas.1602375113},
we ultimately opted for the RPBE+D3 functional in this work.

%\begin{table}%[!htbp]
\begin{table}[htbp]
\caption {The adsorption energies ($E_{ads}$, in eV/H$_2$O) and the vertical distances between Ru-O (in $\AA$) and O-O (in $\AA$) of seven selected water overlayer configurations,
the sublimation energy of ice-Ih ($E_{sub}$,  in eV/H$_2$O), the lattice parameters of bulk Ru (in $\AA$), and the surface energy of Ru(0001) ($E_{sf}$, in eV/$\AA$$^2$)
predicted by PBE+D3, optB88-vdW, SCAN, and RPBE+D3 methods.
Note that the experimental lattice constants of bulk Ru at 0 K are $a$=$b$=2.703~$\AA$ and $c$=4.274~$\AA$~\cite{matthey:/content/journals/10.1595/147106713X665030},
and the experimentally measured sublimation energy of ice-Ih is $-$0.61 eV/H$_2$O~\cite{B808482N}.
}
\begin{tabular}{lccccccccccc}
\hline
 \multicolumn{2}{c}{~} & \multicolumn{3}{c}{\multirow{2}*{$\sqrt{3}$$\times$$\sqrt{3}$}}\qquad\qquad & \multirow{2}*{$\sqrt{3}$$\times$$2\sqrt{3}$}\qquad\qquad & \multicolumn{3}{c}{\multirow{2}*{$2\sqrt{3}$$\times$$2\sqrt{3}$}} & ice-Ih & \multirow{2}*{Lattice} & \multirow{2}*{$E_{sf}$}\\
 \multicolumn{2}{c}{~} & & & & & & & & $E_{sub}$ & & \\
 \cline{3-9}
 \multicolumn{2}{c}{~} & cfg18~\cite{feibelman2002partial} & cfg15~\cite{feibelman2002partial} & cfg42~\cite{PhysRevLett.93.196102}\qquad\qquad & cfg32\cite{PhysRevLett.106.026101}\qquad\qquad & cfg41~\cite{PhysRevLett.93.196102} & cfg25~\cite{gallagher2009order} & cfg28~\cite{gallagher2009order} & (eV/H$_2$O) & parameters& (eV/$\AA$$^2$)\\
 \hline
 \multirow{3}*{PBE+D3} & $E_{ads}$ & -0.72 & -0.72 & -1.02\qquad\qquad & -0.84\qquad\qquad & -0.99 & -0.80 & -0.81 & \multirow{3}*{-0.74} & $a$=2.69 & \multirow{3}*{0.22} \\
 \cline{2-9}
                    & Ru-O & 2.39 & 2.45 & 2.07\qquad\qquad & 2.32\qquad\qquad & 2.08 & 2.32 & 2.29 & & $b$=2.69 & \\
\cline{2-9}
                    & O-O  & 0.75 & 0.54 & 0.05\qquad\qquad & 0.92\qquad\qquad & 0.084 & 0.80 & 0.92 & & $c$=4.28 &\\
 \hline
\multirow{3}*{optB88-vdW} & $E_{ads}$ & -0.70 & -0.67 & -0.98\qquad\qquad & -0.79\qquad\qquad & -0.96 & -0.74 & -0.76 & \multirow{3}*{-0.71} & $a$=2.72 & \multirow{3}*{0.18} \\
 \cline{2-9}
                    & Ru-O & 2.45 & 2.51 & 2.08\qquad\qquad & 2.34\qquad\qquad & 2.10 & 2.35 & 2.31 & & $b$=2.72 &\\
\cline{2-9}
                    & O-O  & 0.79 & 0.54 & 0.05\qquad\qquad & 0.92\qquad\qquad & 0.087 & 0.81 & 0.94 & & $c$=4.31 &\\
 \hline
\multirow{3}*{SCAN} & $E_{ads}$ & -0.68 & -0.65 & -1.01\qquad\qquad & -0.79\qquad\qquad & -0.99 & -0.73 & -0.75  & \multirow{3}*{-0.72} & $a$=2.69 & \multirow{3}*{0.19}\\
 \cline{2-9}
                    & Ru-O & 2.47 & 2.55 & 2.06\qquad\qquad & 2.33\qquad\qquad & 2.07 & 2.33 & 2.27 & & $b$=2.69 &\\
\cline{2-9}
                    & O-O  & 0.83 & 0.47 & 0.05\qquad\qquad & 1.00\qquad\qquad & 0.12 & 0.86 & 1.00 & & $c$=4.26 &\\
 \hline
\multirow{3}*{RPBE+D3} & $E_{ads}$ & -0.60 & -0.58 & -0.83\qquad\qquad & -0.69\qquad\qquad & -0.82 & -0.66 & -0.67  & \multirow{3}*{-0.62} & $a$=2.69 & \multirow{3}*{0.22}\\
 \cline{2-9}
                    & Ru-O & 2.49 & 2.60 & 2.08\qquad\qquad & 2.37\qquad\qquad & 2.11 & 2.39 & 2.33 & & $b$=2.69 &\\
\cline{2-9}
                    & O-O  & 0.81 & 0.47 & 0.06\qquad\qquad & 0.90\qquad\qquad & 0.09 & 0.83 & 0.97 & & $c$=4.27 &\\
 \hline
\end{tabular}
\label{tables1}
\end{table}

\newpage
\clearpage
%--------------------------------------------------------------------------------
\section{Training and validation of moment tensor potential}\label{sec:training_details}
%--------------------------------------------------------------------------------

The training dataset construction followed the scheme outlined in our previous works~\cite{PhysRevLett.130.078001,Liu2024}.
Specifically, the training structures were initially sampled using an on-the-fly active learning scheme
based on the Bayesian linear regression~\cite{JinnouchiPRL2019,JinnouchiPRB2019}.
This enabled us to efficiently sample the phase space and automatically collect the representative
training structures during the molecular dynamics (MD) simulations.
The cutoff radius for the three-body descriptors and the width of the Gaussian functions
used for broadening the atomic distributions of the three-body descriptors were set to 6~$\AA$~and 0.5~$\AA$, respectively.
The number of radial basis functions used to expand the radial descriptor was set to 10.
Then, the training dataset was expanded iteratively using the active learning method implemented in the MLIP package~\cite{Novikov_2021}.
For the training of the moment tensor potential (MTP), a cutoff radius of 6.0 $\AA$ was used, and the number of radial basis was set to be 8.
The MD samplings were conducted by heating the water overlayers at various coverages from 10 K to 400 K using the NVT canonical ensemble.
Additionally, configurations derived from path-integral molecular dynamics simulations were incorporated into the training dataset, guided by the active learning framework~\cite{Novikov_2021}.

\begin{figure}[!htbp]
\begin{center}
\includegraphics[width=1.0\textwidth,trim = {0.0cm 0.0cm 0.0cm 0.0cm}, clip]{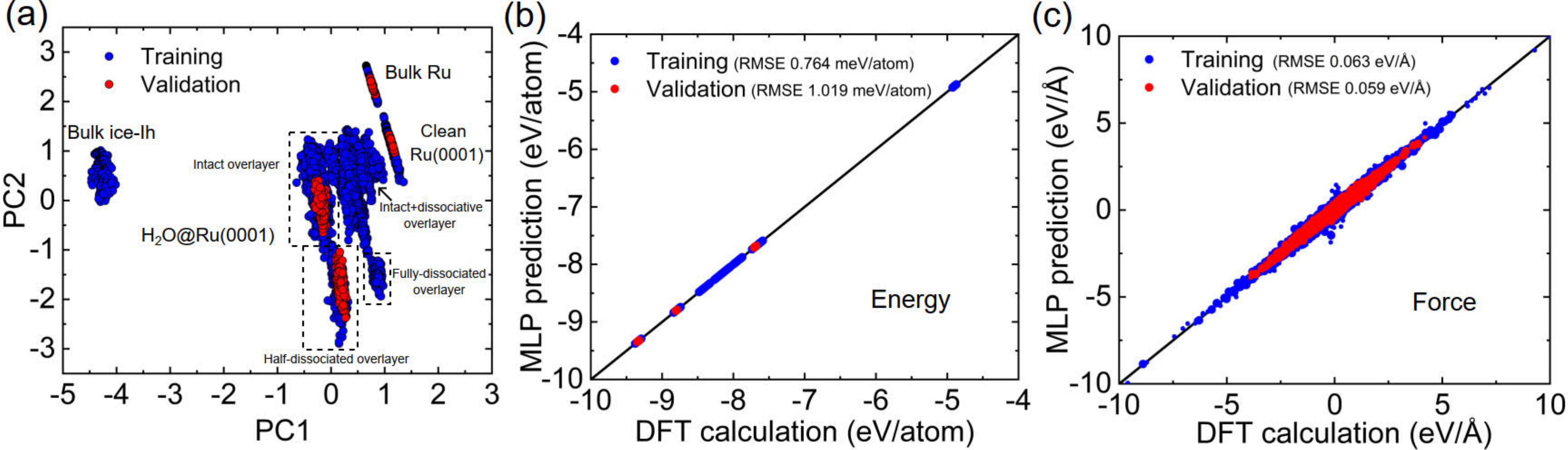}
\end{center}
\caption{(a) Kernel principal component analysis map of training structures (blue circles) and validation structures (red circles).
(b) Machine-learning potential (MLP) predicted energies versus DFT results. (c) MLP predicted forces versus DFT results.
}
\label{figs3_MLP_DFT}
\end{figure}

The eventual training dataset contained 6,813 structures
including water molecule, bulk ice-Ih, bulk Ru, and clean Ru(0001) surface,
as well as intact and dissociated water overlayers at various coverages up to 2/3 monolayer (ML) (see Table~\ref{tab:training_dataset}).
To enhance efficiency, the ultimate MTP was fitted with the optimized basis sets
by refining the contraction process of moment tensors using our in-house code (IMR-MLP)~\cite{wang2024}.
The MTP was validated on a test dataset containing 600 structures
including bulk Ru, clean Ru(0001) surface,
intact and half-dissociated water overlayers at 2/3 ML (see Table~\ref{tab:test_dataset}).
These test structures were randomly sampled from additional MD simulations.
The kernel principal component analysis on both the training and validation datasets were presented in Fig.~\ref{figs3_MLP_DFT}(a).
The root-mean-square errors in energies and forces for the training and validation datasets were displayed  in Table~\ref{tables4},
while the parity plots for MTP-predicted energies and forces against DFT results were depicted in Fig.~\ref{figs3_MLP_DFT}(b) and Fig.~\ref{figs3_MLP_DFT}(c),
showcasing the high accuracy of the generated MTP.

\begin{table}[!htbp]
\caption {Summary of the structures included in the training dataset.}
\begin{ruledtabular}
\begin{tabular}{crr}
 H$_2$O coverages (ML) & Structure type                  & Number of structures \\
\hline
 ---   &  Bulk ice-Ih of 36-atom cell                    &  306  \\
 ---   &  Bulk Ru of 54-atom cell                        &  111  \\
 0/3   &  Clean $\sqrt{3}$$\times$$\sqrt{3}$ Ru(0001)    &  183  \\
 0/6   &  Clean $\sqrt{3}$$\times$2$\sqrt{3}$ Ru(0001)   &  287  \\
 0/12  &  Clean 2$\sqrt{3}$$\times$2$\sqrt{3}$ Ru(0001)  &  176  \\
 2/3   &  2H$_2$O@$\sqrt{3}$$\times$$\sqrt{3}$ Ru(0001)  &  1329 \\
 4/6   &  4H$_2$O@$\sqrt{3}$$\times$2$\sqrt{3}$ Ru(0001) &  371  \\
 1/9   &  1H$_2$O@3$\times$3 Ru(0001)                    &  28   \\
 2/9  &  2H$_2$O@3$\times$3 Ru(0001)&  81\\
 3/9  &  3H$_2$O@3$\times$3 Ru(0001)&  89\\
 4/9  &  4H$_2$O@3$\times$3 Ru(0001)&  106\\
 6/9  &  6H$_2$O@3$\times$3 Ru(0001)&  679\\
 7/9  &  7H$_2$O@3$\times$3 Ru(0001)&  120\\
 8/12  &  8H$_2$O@2$\sqrt{3}$$\times$2$\sqrt{3}$ Ru(0001) &  684\\
 8/12  &  8H$_2$O@3$\times$2$\sqrt{3}$ Ru(0001)&  1439\\
 10/15  &  10H$_2$O@3$\times$$\sqrt{21}$ Ru(0001) &  27 \\
 12/18  &  12H$_2$O@3$\times$3$\sqrt{3}$ Ru(0001) &  226\\
 4/22  &  4H$_2$O@11$\times$$\sqrt{3}$ Ru(0001) &   2 \\
 5/22  &  5H$_2$O@11$\times$$\sqrt{3}$ Ru(0001) &   62 \\
6/22  &   6H$_2$O@11$\times$$\sqrt{3}$ Ru(0001) &   15 \\
7/22  &   7H$_2$O@11$\times$$\sqrt{3}$ Ru(0001) &  41\\
8/22  &   8H$_2$O@11$\times$$\sqrt{3}$ Ru(0001) &   150 \\
8/24  &   8H$_2$O@3$\times$4$\sqrt{3}$ Ru(0001) &  121 \\
10/24  &   10H$_2$O@3$\times$4$\sqrt{3}$ Ru(0001)&   117\\
16/24  &   16H$_2$O@3$\times$4$\sqrt{3}$ Ru(0001) &  28 \\
12/32  &   12H$_2$O@8$\times$2$\sqrt{3}$ Ru(0001)&   4\\
24/36  & 24H$_2$O@6$\times$3$\sqrt{3}$ Ru(0001) &  21 \\
26/40  &  26H$_2$O@$\sqrt{37}$$\times$$\sqrt{37}$ Ru(0001)&   10\\
\hline
Total    &                                    & 6813\\
\end{tabular}
\end{ruledtabular}
\label{tab:training_dataset}
\end{table}

\begin{table}[!htbp]
\caption {Summary of the structures included in the validation dataset.}
\begin{ruledtabular}
\begin{tabular}{crr}
H$_2$O coverages (ML)  & Structure type                  & Number of structures \\
\hline
---  &  Bulk Ru of 54-atom cell                              &   100  \\
0/12 &  Clean 2$\sqrt{3}$$\times$2$\sqrt{3}$ Ru(0001)        &   100  \\
8/12  &  8H$_2$O@2$\sqrt{3}$$\times$2$\sqrt{3}$ Ru(0001)     &  400  \\
\hline
Total    &                                    & 600  \\
\end{tabular}
\end{ruledtabular}
\label{tab:test_dataset}
\end{table}

\begin{table}[!htbp]
\caption {The root-mean-square errors (RMSEs) in energies per atom (meV/atom) and forces (eV/$\AA$)
for the training and validation datasets.
}
\begin{ruledtabular}
\begin{tabular}{ccc}
 & Training dataset (6,813 structures) & Validation dataset (600 structures)  \\
 \hline
Energy              & 0.764 & 1.019 \\
Force               & 0.063 & 0.059  \\
\end{tabular}
\end{ruledtabular}
\label{tables4}
\end{table}

\newpage
\clearpage
%--------------------------------------------------------------------------------
\section{Examining disorder effects}\label{sec:disorder}
%--------------------------------------------------------------------------------

\begin{figure}[!htbp]
\begin{center}
\includegraphics[width=0.9\textwidth,trim = {0.0cm 0.0cm 0.0cm 0.0cm}, clip]{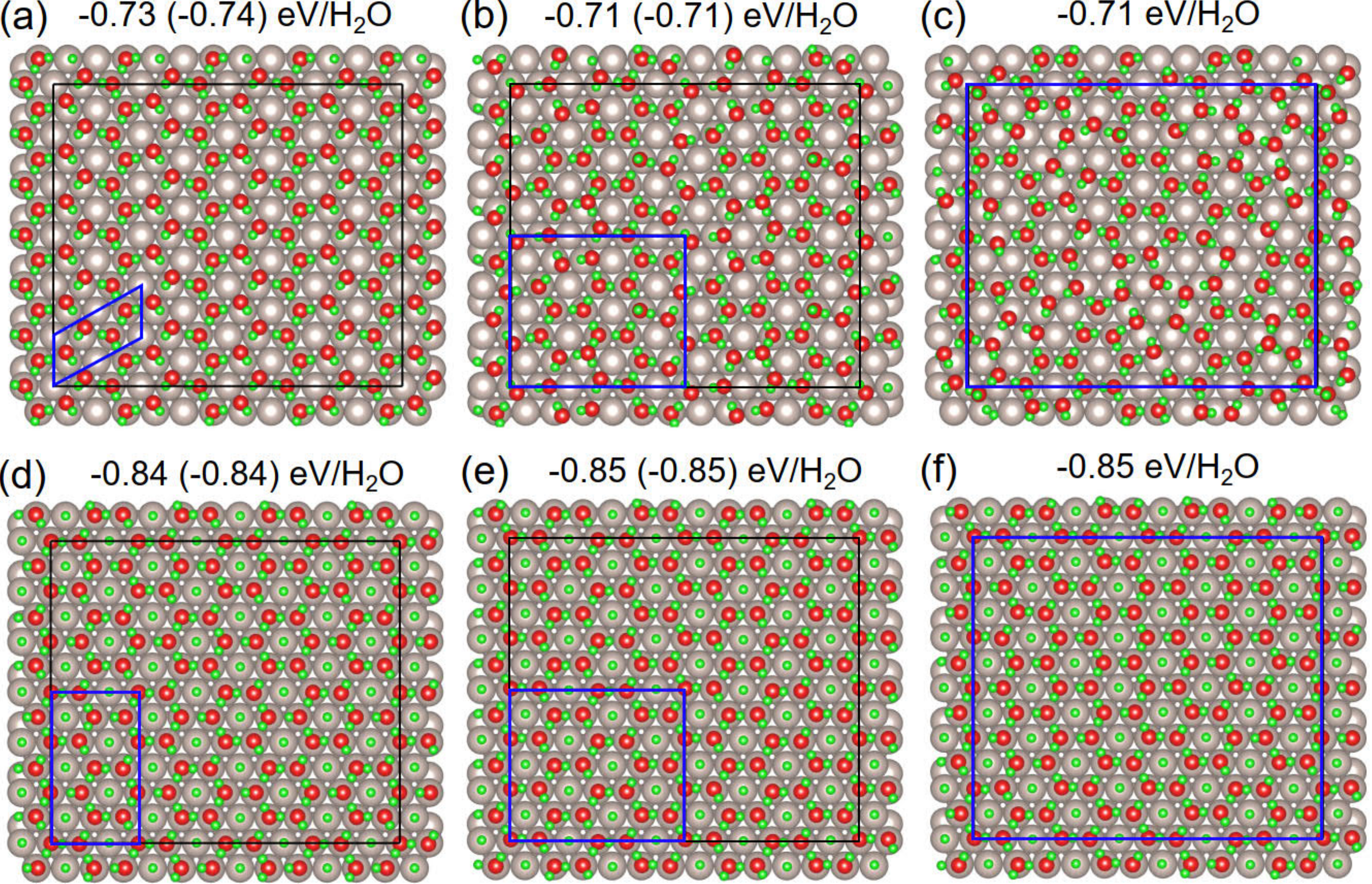}
\end{center}
\caption{(a) The extended chains model with a unit cell of $\sqrt{3}{\times}2\sqrt{3}$ (i.e., cfg 40~\cite{PhysRevLett.106.026101}).
(b) The extended chains model with a unit cell of $6\times3\sqrt{3}$.
(c) The extended chains model with a unit cell of $12\times6\sqrt{3}$.
(d) The half-dissociated model with a unit cell of $3\sqrt{3}\times3$ (i.e., cfg 43~\cite{feibelman2002partial}).
(e) The half-dissociated model with a unit cell of $6\times3\sqrt{3}$.
(f) The half-dissociated model with a unit cell of $12\times6\sqrt{3}$.
The blue lines indicate the unit cell.
Note that (b), (c), (e), and (f) depict the lowest-energy configurations,
each of which was identified from 90  independent simulated annealing MD simulations followed by structural relaxations.
All the configurations are expanded to the $12\times6\sqrt{3}$ supercell for aiding the visibility.
The MTP and DFT (in brackets) predicted adsorption energies (in eV/H$_2$O) are indicated above each configuration.
}
\label{figs4_heat_cool}
\end{figure}

\newpage
\clearpage
%--------------------------------------------------------------------------------
\section{Energy barrier for water dissociation at 0 K}\label{sec:NEB}
%--------------------------------------------------------------------------------

\begin{figure}[!htbp]
\begin{center}
\includegraphics[width=0.9\textwidth,trim = {0.0cm 0.0cm 0.0cm 0.0cm}, clip]{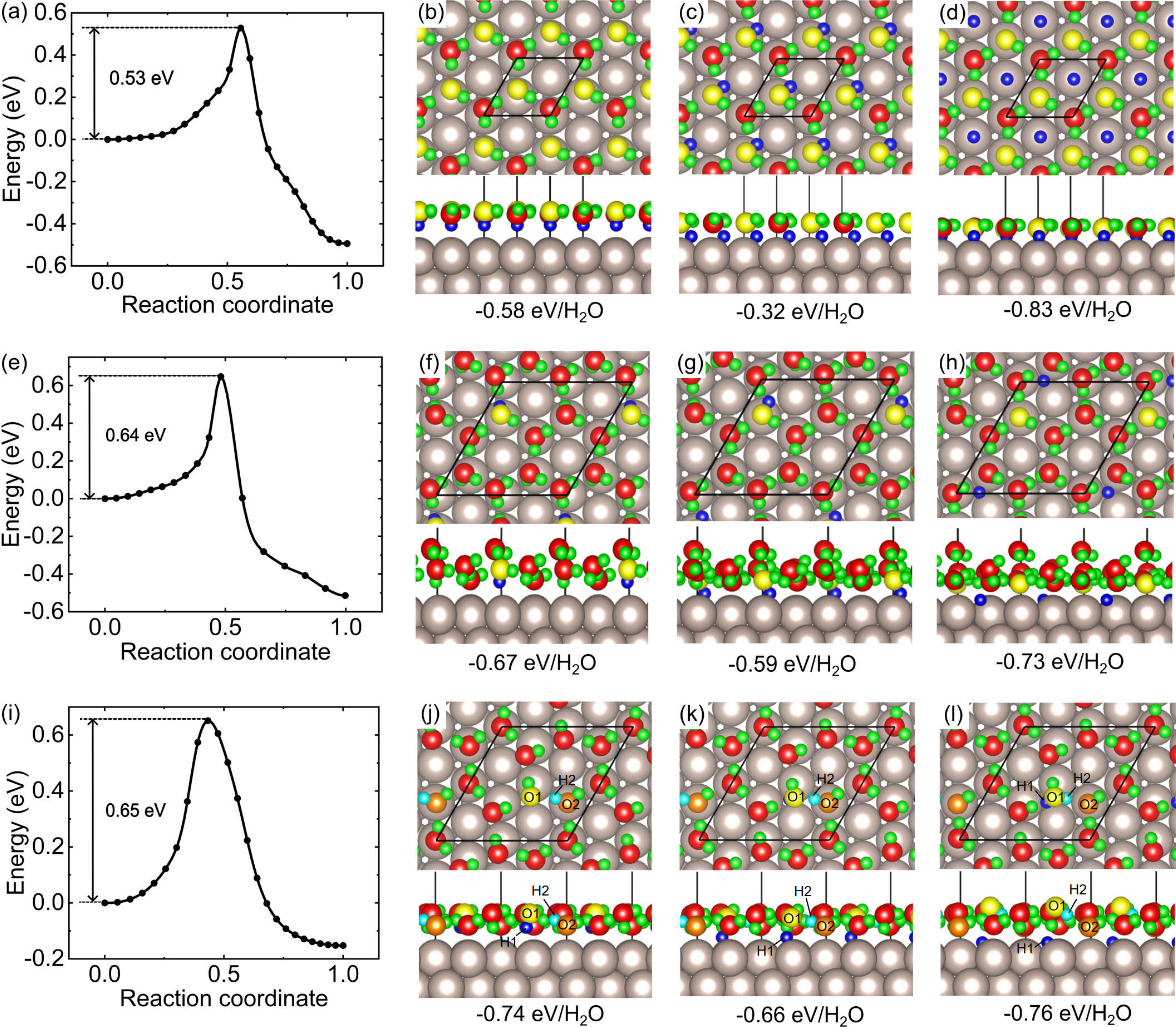}
\end{center}
\caption{(a) The MTP-predicted energy profile for the dissociation of half the water in the H-down bilayer structure
using the climbing image nudged elastic band method~\cite{Henkelman2000ACI}.
Panels (b), (c), and (d) depict the initial, transitional, and final configurations, respectively, corresponding to the dissociation process in panel (a).
(e) The MTP-predicted energy profile for the dissociation of 1/8 of the water in the H-down bilayer structure.
Panels (f), (g), and (h) depict the initial, transitional, and final configurations, respectively, corresponding to the dissociation process in panel (e).
(i) The MTP-predicted energy profile for the dissociation of 1/8 of the water in the extended chain structure.
Panels (j), (k), and (l) depict the initial, transitional, and final configurations, respectively, corresponding to the dissociation process in panel (i).
Note that the brown spheres represent Ru atoms,
while the red and yellow spheres stand for oxygen atoms of intact and dissociated water molecules, respectively.
Moreover, the green and blue spheres denote the hydrogen atoms in intact water molecules and the dissociative hydrogen atom, respectively.
In panels (j), (k), and (l), the cyan sphere indicates the hydrogen atom (referred to as ``H2") participating in an indirect dissociation process.
Specifically, two H$_2$O molecules (labeled O1 and O2) are involved in the dissociation process (i).
The MTP predicted adsorption energies (in eV/H$_2$O) are indicated below each configuration.
}
\label{figs5_barrier}
\end{figure}

\newpage
\clearpage
%--------------------------------------------------------------------------------
\section{Energy barrier for water desorption at 0 K}\label{sec:NEB_desorption}
%--------------------------------------------------------------------------------

\begin{figure}[!htbp]
\begin{center}
\includegraphics[width=0.9\textwidth,trim = {0.0cm 0.0cm 0.0cm 0.0cm}, clip]{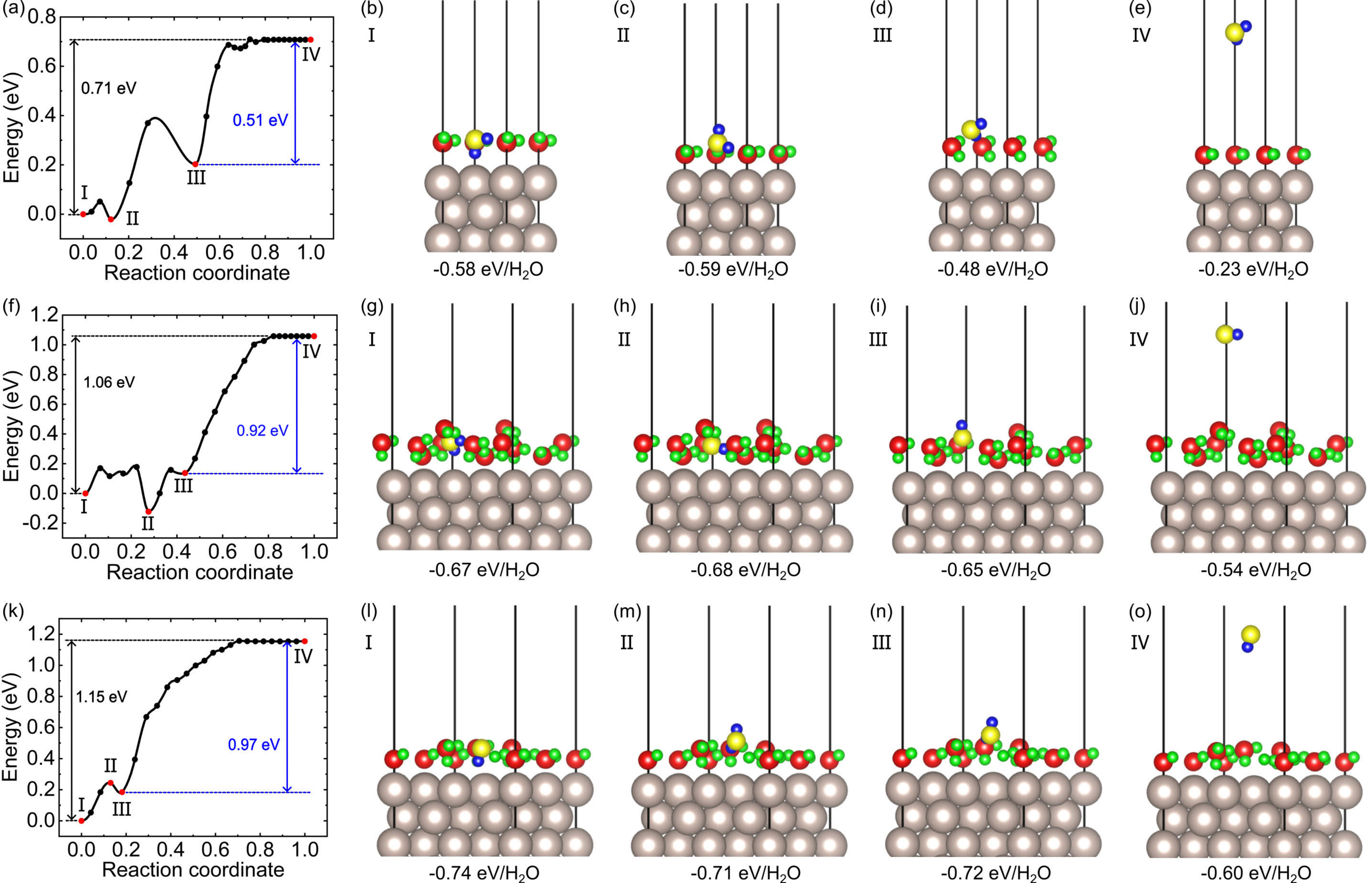}
\end{center}
\caption{(a) The MTP-predicted energy profile for the desorption of half the water in the H-down bilayer structure
using the climbing image nudged elastic band method~\cite{Henkelman2000ACI}.
Panels (b), (c), (d), and (e) depict the configurations marked in panel (a).
(f) The MTP-predicted energy profile for the desorption of 1/8 of the water in the H-down bilayer structure.
Panels (g), (h), (i), and (j) depict the configurations marked in panel (f).
(k) The MTP-predicted energy profile for the desorption of 1/8 of the water in the extended chain structure.
Panels (l), (m), (n), and (o) depict the configurations marked in panel (k).
In panels (a), (f), and (k), the black values indicate desorption energies,
while the blue values represent desorption energy barriers for the rate-determining step.
Note that the H-down bilayer structure is a metastable configuration as compared to the extended chain structure.
The atomic representations are as follows: brown spheres for Ru atoms, red and yellow spheres for oxygen atoms in intact and desorbed water molecules, respectively, and green and blue spheres for hydrogen atoms in intact and desorbed water molecules, respectively. Below each configuration, the MTP-predicted adsorption energies are provided.
}
\label{figs6_desorption_barrier}
\end{figure}

\newpage
\clearpage
%--------------------------------------------------------------------------------
\section{Path-integral and classical molecular dynamics simulations}\label{sec:MD_details}
%--------------------------------------------------------------------------------

\begin{figure}[!htbp]
\begin{center}
\includegraphics[width=0.6\textwidth,trim = {0.0cm 0.0cm 0.0cm 0.0cm}, clip]{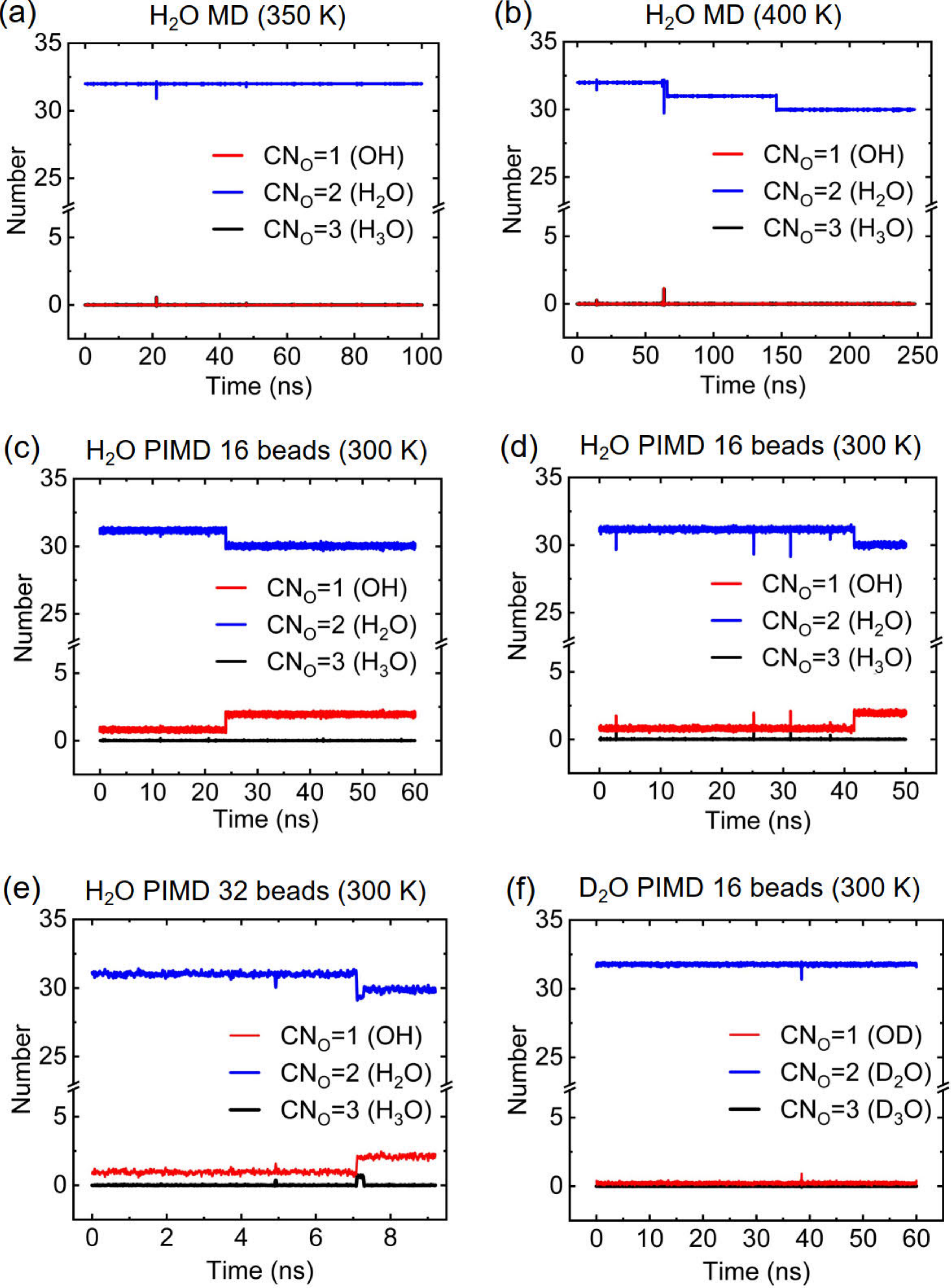}
\end{center}
\caption{Time evolution of the number of OH (CN$_{\rm O}$=1),  H$_2$O (CN$_{\rm O}$=2), and H$_3$O (CN$_{\rm O}$=3) species
in (a) a classical MD simulation at 350 K, (b) a classical MD simulation at 400 K, and (c)-(d) two independent PIMD simulations at 300 K with 16 beads.
CN$_{\rm O}$ represents the coordination number of the oxygen atom.
Note that the reduction of H$_2$O in panel (b)  is due to the evaporation of water at 400 K into the vacuum. However, even at this high temperature, no dissociation of water is observed.
(e) Time evolution of the oxygen atom's coordination number in a PIMD simulation of H$_2$O adsorption at 300 K with 32 beads.
(f) Time evolution of the oxygen atom's coordination number in a PIMD simulation of D$_2$O adsorption at 300 K with 16 beads.
Here, all simulations employed the same initial supercell structure as used in Fig.~2 of the main text.
}
\label{figs7_md_pimd}
\end{figure}

\begin{figure}[!htbp]
\begin{center}
\includegraphics[width=0.7\textwidth,trim = {0.0cm 0.0cm 0.0cm 0.0cm}, clip]{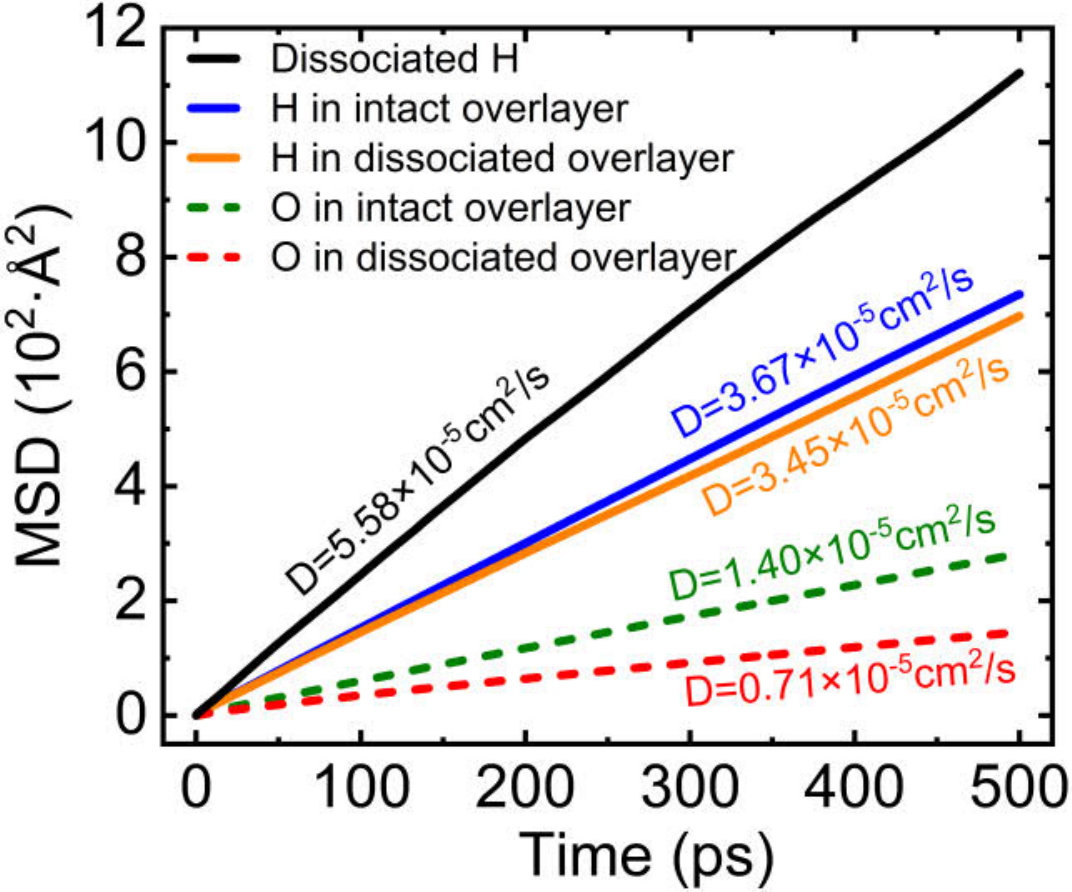}
\end{center}
\caption{Time evolution of the mean square displacements (MSD) for dissociated H atoms,
as well as H and O atoms within intact or dissociated water overlayers predicted by the PIMD simulations at 300 K.
The diffusion coefficients are indicated as calculated using the formula $D=\frac{1}{4}\frac{d{\rm MSD}}{dt}$ where $t$ represents the time.
}
\label{figs8_MSD}
\end{figure}

\begin{figure}[!htbp]
\begin{center}
\includegraphics[width=0.7\textwidth,trim = {0.0cm 0.0cm 0.0cm 0.0cm}, clip]{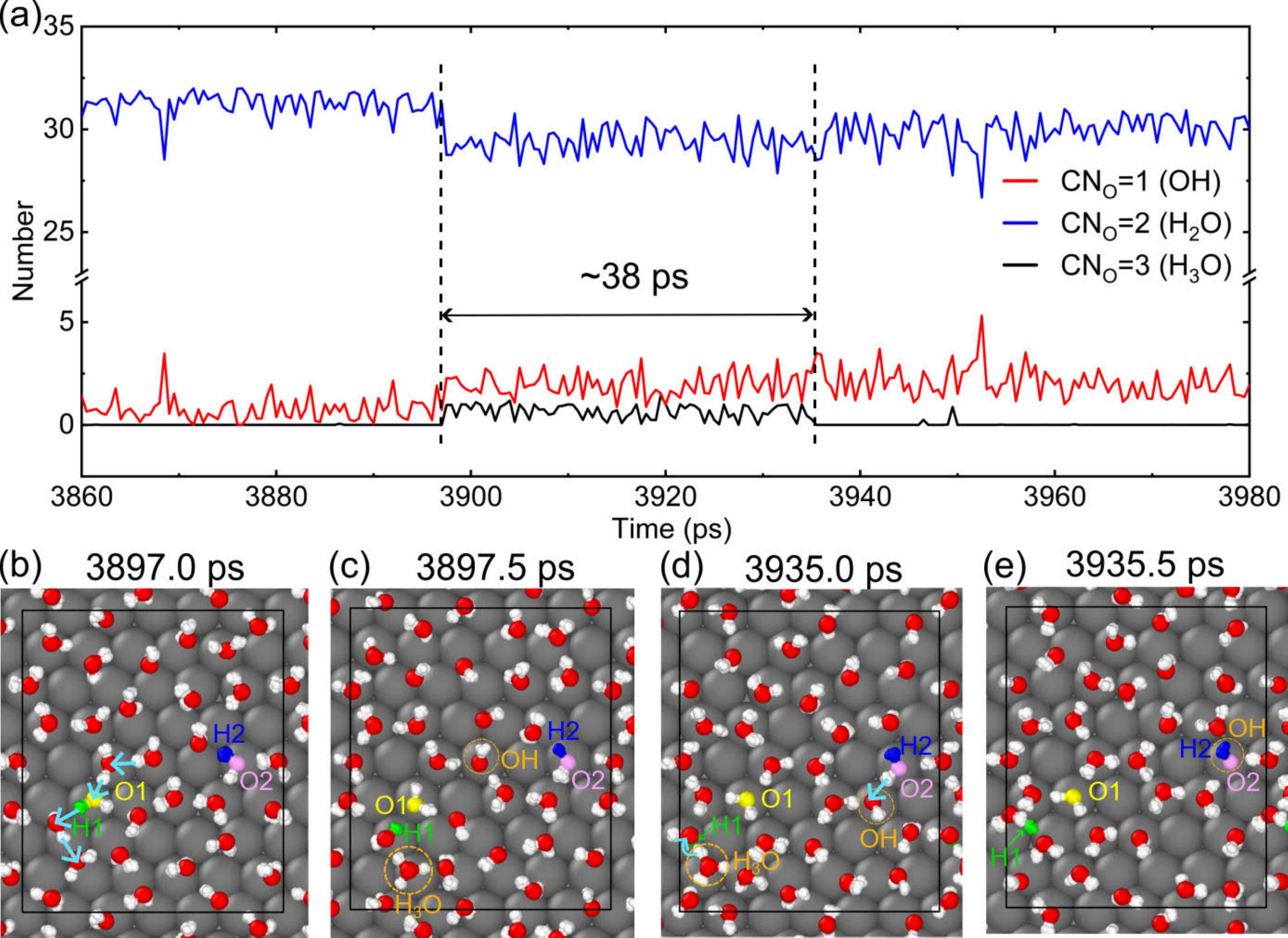}
\end{center}
\caption{(a) Time evolution of the number of OH (CN$_{\rm O}$=1),  H$_2$O (CN$_{\rm O}$=2), and H$_3$O (CN$_{\rm O}$=3) species
in the same PIMD simulation as in Fig.~2(b) of the main text, but with a focus on the time period surrounding the first dissociation event.
CN$_{\rm O}$ represents the coordination number of the oxygen atom.
The estimated lifetime of the H$_3$O species is indicated.
(b)-(e) Snapshots from the PIMD trajectory. The tracked O and H atoms as well as the H$_3$O and OH species are highlighted.
The cyan arrows depict the proton transfer process.
}
\label{figs9_pimd_1st}
\end{figure}

\begin{figure}[!htbp]
\begin{center}
\includegraphics[width=0.7\textwidth,trim = {0.0cm 0.0cm 0.0cm 0.0cm}, clip]{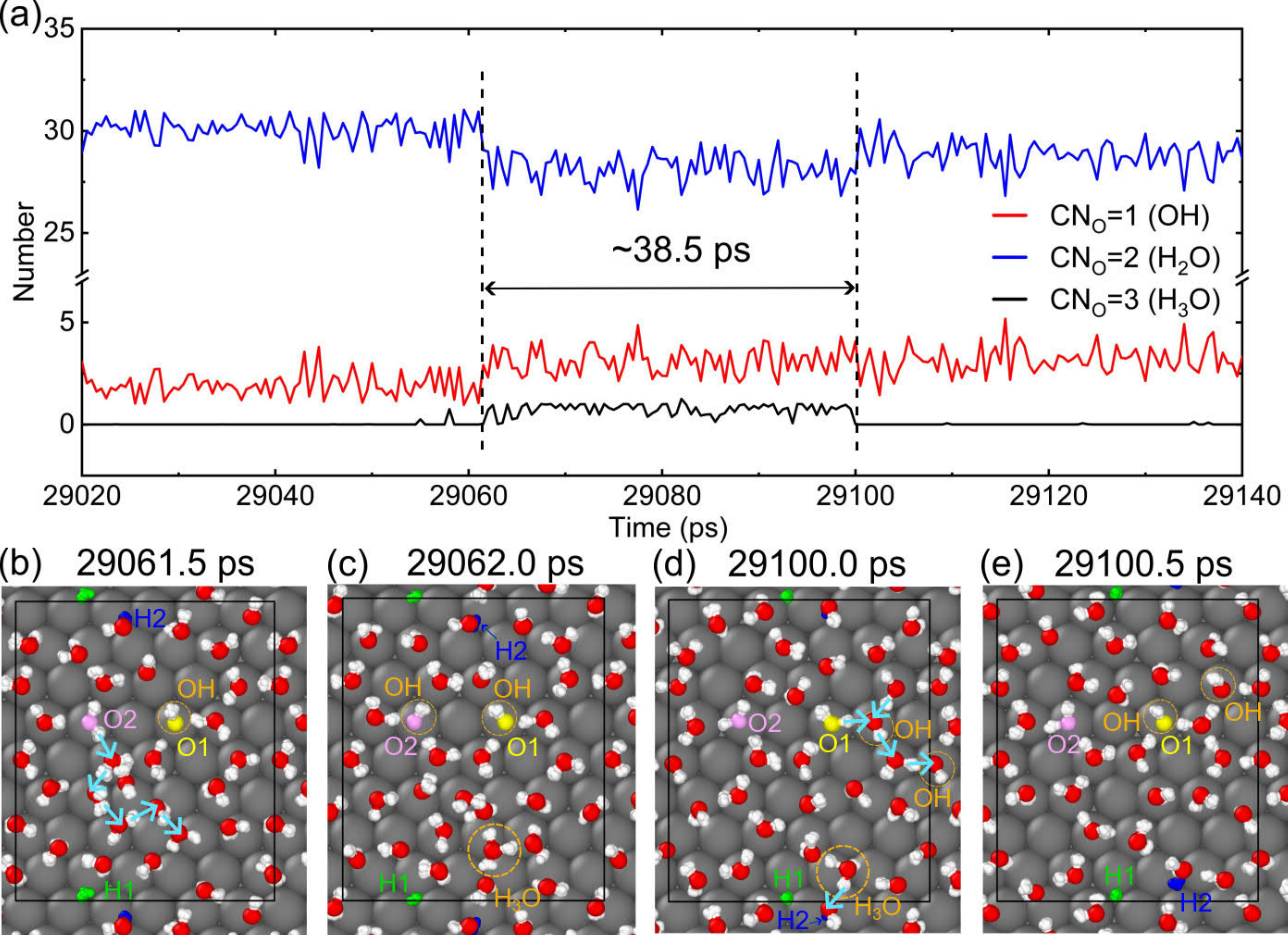}
\end{center}
\caption{Same as Fig.~\ref{figs9_pimd_1st}, but with a focus on the time period surrounding the second dissociation event.
}
\label{figs10_pimd_2nd}
\end{figure}

\begin{figure}[!htbp]
\begin{center}
\includegraphics[width=0.7\textwidth,trim = {0.0cm 0.0cm 0.0cm 0.0cm}, clip]{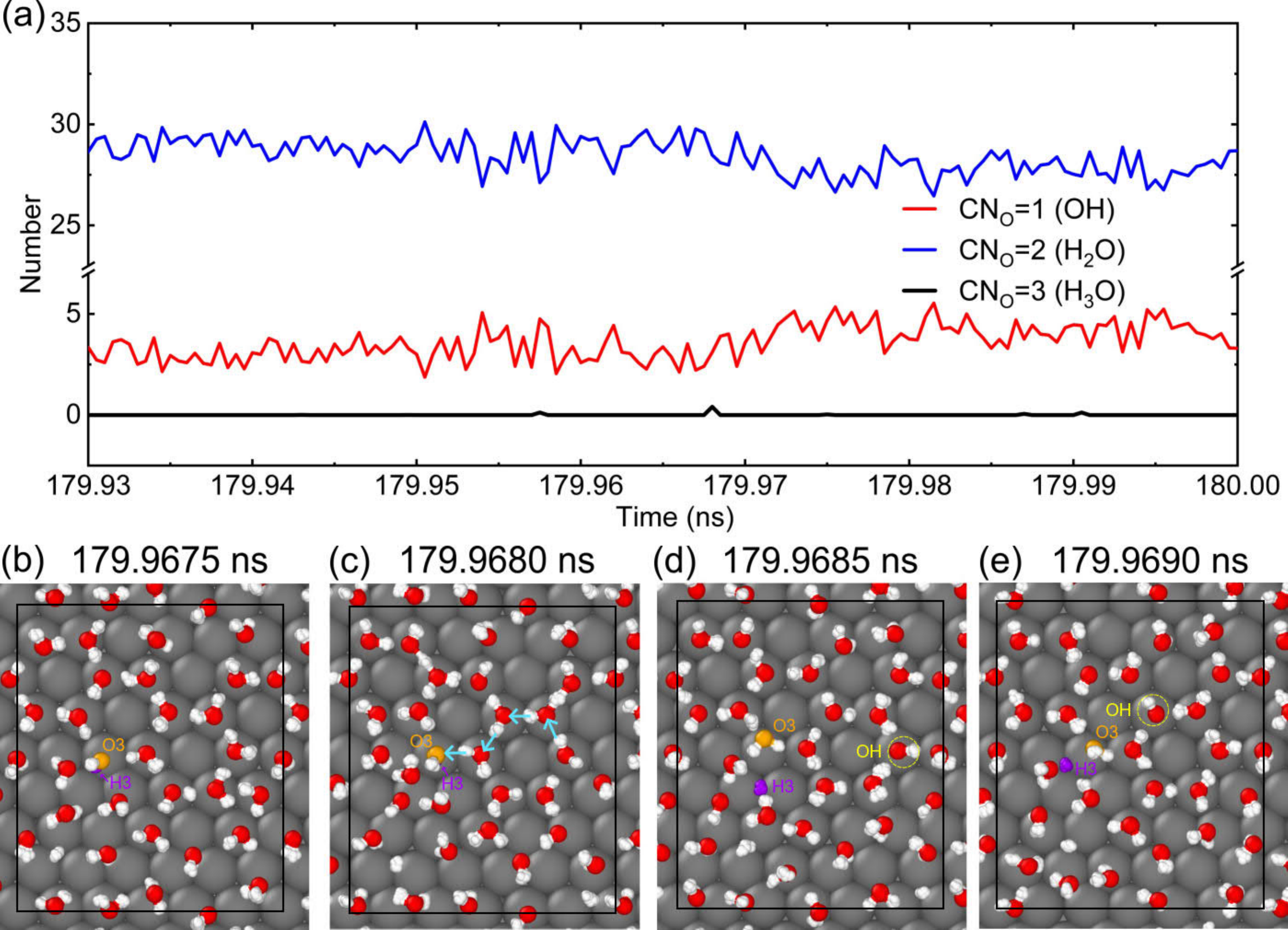}
\end{center}
\caption{Same as Fig.~\ref{figs9_pimd_1st}, but with a focus on the time period surrounding the third dissociation event.
}
\label{figs11_pimd_3rd}
\end{figure}

The classical molecular dynamics (MD) and path-integral molecular dynamics (PIMD) simulations
were conducted using the LAMMPS code~\cite{Thompson2022} in the NVT canonical ensemble.
Unless explicitly stated otherwise, all simulations were conducted at room temperature (300 K).
We note that, the relatively high simulated temperature we employed was intended to accelerate the dynamics of the simulation.
The temperature was controlled using a Nos\'e-Hoover thermostat~\cite{Nose1991}.
The time step was set to 0.5 fs.
In the PIMD simulations, each nucleus was represented by 16 beads and the statistical analyses were derived by averaging these 16 beads.
Note that we also conducted the PIMD simulations employing 32 beads and the similar water dissociation was observed [see Fig.~\ref{figs7_md_pimd}(e)].
A $6\times4\sqrt{3}$ supercell of the most stable intact overlayer configuration (i.e., extended chains model) was employed as the initial structure.
Figs.~\ref{figs7_md_pimd}(a)-(e) illustrate the time evolution of the number of OH,  H$_2$O,
and H$_3$O species in a series of independent MD and PIMD simulations.
Fig.~\ref{figs7_md_pimd}(f) shows the time evolution of the oxygen atom's coordination number in a PIMD simulation of D$_2$O adsorption at 300 K with 16 beads,
where no D$_2$O dissociation is observed, highlighting the pronounced kinetic isotope effect.
Figure~\ref{figs8_MSD} displays the time evolution of the mean square displacements predicted by MTP
for the dissociated H atoms, the H and O atoms within the intact or dissociated water overlayers.
Figure~\ref{figs9_pimd_1st} zooms in on Fig.~2(b) from the main text, but focuses on the time period surrounding the first dissociation event.
Figure~\ref{figs10_pimd_2nd} is similar to Fig.~\ref{figs9_pimd_1st}, but focuses on the time period surrounding the second dissociation event.
Figure~\ref{figs11_pimd_3rd} is similar to Fig.~\ref{figs9_pimd_1st}, but focuses on the time period surrounding the third dissociation event.

\begin{figure}[!htbp]
\begin{center}
\includegraphics[width=0.8\textwidth,trim = {0.0cm 0.0cm 0.0cm 0.0cm}, clip]{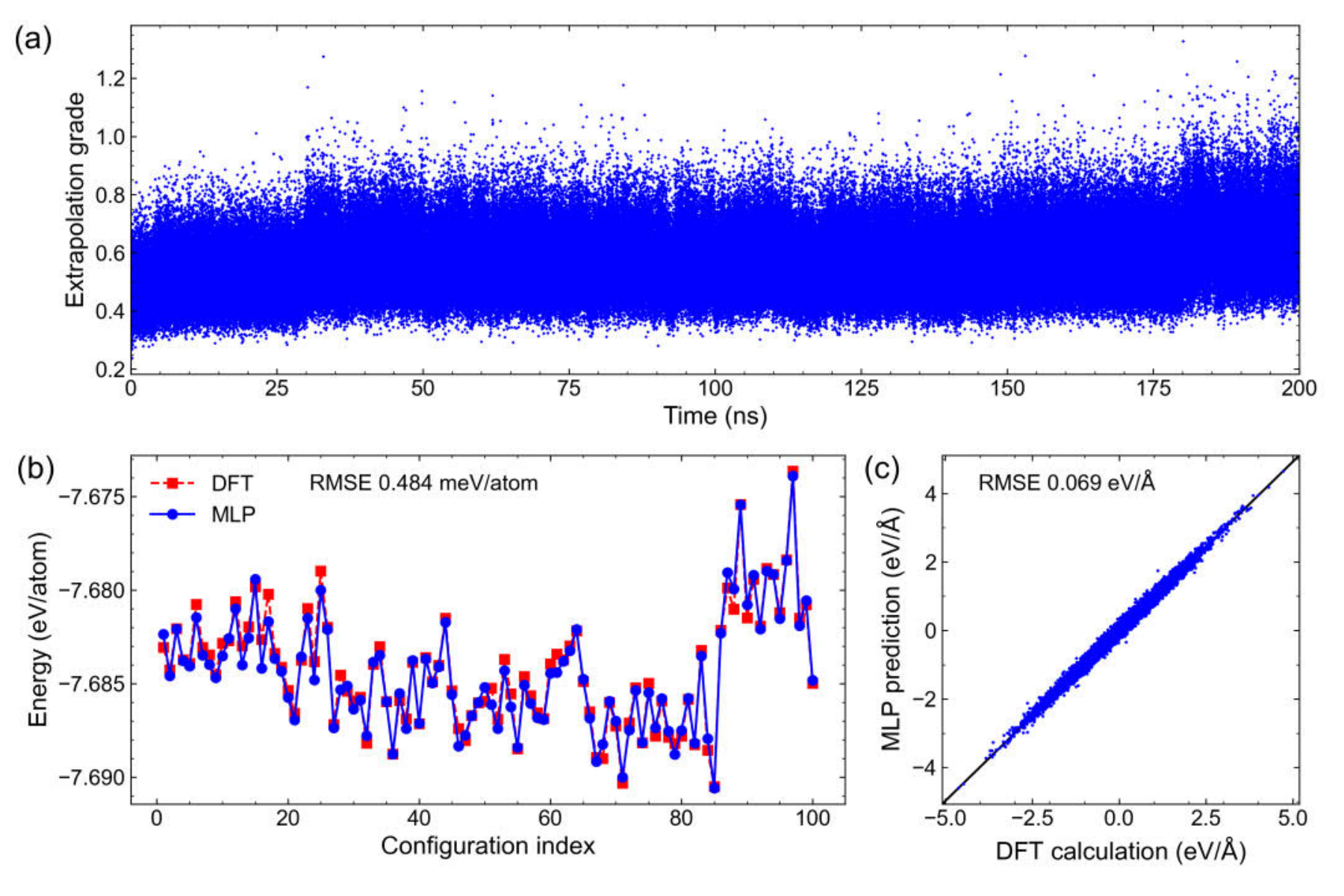}
\end{center}
\caption{(a) Time evolution of the extrapolation grade for all centroid structures obtained
by averaging all beads in the 200-nanosecond PIMD simulations  (corresponding to the case in Fig.~2(b) of the main text).
(b) Comparison of energies predicted by MLP and DFT for 100 randomly selected centroid structures.
(c) MLP-predicted forces versus DFT results for the same 100 centroid structures.
The root-mean-square errors (RMSEs) for energies and forces are also provided for quantitative assessment.
}
\label{figs12_extrapolation_grade}
\end{figure}

We emphasize that the predictions derived from our MLP are both accurate and robust. This reliability stems from the generation of our MLP using a highly descriptive training dataset that comprehensively spans the relevant phase space. To validate this, we computed the extrapolation grade for centroid structures obtained by averaging all beads from the long-timescale PIMD trajectories corresponding to Fig.~2(b) in the main text. As illustrated in Fig.~\ref{figs12_extrapolation_grade}(a), all centroid structures exhibit an extrapolation grade below 1.3, confirming that our MLP operates within the interpolation regime~\cite{Novikov_2021}. This result underscores the reliability of our MLP-derived conclusions. To further validate the MLP, we randomly selected 100 centroid structures from the PIMD trajectories, encompassing both intact and dissociated water configurations. We compared the energies and forces predicted by the MLP with those computed using DFT. As demonstrated in Figs.~\ref{figs12_extrapolation_grade}(b) and (c), the MLP predictions show excellent agreement with the DFT results, reinforcing the accuracy and reliability of our MLP-accelerated PIMD predictions.

%--------------------------------------------------------------------------------
\section{Data availability}
%--------------------------------------------------------------------------------

The data supporting the findings of this study, such as the training and validation datasets, the developed MTP, and the initial and final structures as well as the LAMMPS input scripts for both MD and PIMD simulations, are publicly available at https://github.com/wftao1995/Water-Dissociation-on-Ru-0001-Dataset.

\bibliography{supplementReference} %here in windows there is no surfix .bib